\documentclass[11pt]{article}

\usepackage{latexsym,amsfonts,amsmath,amssymb,wasysym,enumerate,epsfig}
\usepackage{theorem}
\usepackage{mathrsfs}
\usepackage{tikz}
\usepackage{ulem}
\usepackage{bm}
\usepackage[all]{xy}

\usepackage{color}
\usepackage{hyperref}

\numberwithin{equation}{section}
\newtheorem{definition}{Definition}[section]

\newtheorem{remark}[definition]{Remark}

\theorembodyfont{\rmfamily}

\def\cA{{\cal A}}    \def\cB{{\cal B}}    \def\cC{{\cal C}}

    \def\cK{{\cal K}}    
        
\def\cP{{\cal P}}        \def\cR{{\cal R}}
\def\cS{{\cal S}}        \def\cU{{\cal U}}
\def\cV{{\cal V}}         
\def\cY{{\cal Y}}    

\newcommand{\CC}{{\mathbb C}}
\newcommand{\UU}{{\mathbb U}}

\def\and{\quad\mbox{and}\quad}

\newcommand\bra[1]{{\langle#1|}}
\newcommand\ket[1]{{|#1\rangle}}

\newcommand\bbra[1]{{\langle\!\langle#1|}}
\newcommand\kket[1]{{|#1\rangle\!\rangle}}

\newcommand{\llangle}{\langle\!\langle}
\newcommand{\rrangle}{\rangle\!\rangle}
\newcommand{\steady}{|{\cal S}\rangle} 
\advance\textheight by \topskip
\begin{document}
\setcounter{page}{0}
\pagestyle{empty}
%
%
\begin{center}

 {\LARGE  {\sffamily  Integrable Floquet dynamics, generalized exclusion processes and 
 ``fused'' matrix ansatz} }\\[1cm]

\vspace{10mm}
  
{\Large 
 M. Vanicat$^{a}$\footnote{matthieu.vanicat@fmf.uni-lj.si}}\\[.41cm] 
{\large $^{a}$  Faculty of Mathematics and Physics, University of Ljubljana,\\[.242cm]
 Jadranska 19, SI-1000 Ljubljana, Slovenia. }
\end{center}
\vfill

\begin{abstract}
We present a general method for constructing integrable stochastic processes, with two-step discrete time Floquet dynamics, 
from the transfer matrix formalism. The models can be interpreted as a discrete time parallel update.
The method can be applied for both periodic and open boundary conditions. We also show how the stationary distribution can be built as a matrix 
product state. As an illustration we construct parallel discrete time dynamics 
associated with the R-matrix of the SSEP and of the ASEP, and provide the associated stationary distributions in a matrix product form.
We use this general framework to introduce new integrable generalized exclusion processes, where a fixed number of particles is allowed on 
each lattice site in opposition to the (single particle) exclusion process models. 
They are constructed using the fusion procedure of R-matrices (and K-matrices for open boundary conditions)
for the SSEP and ASEP. We develop a new method, that we named ``fused'' matrix ansatz, to build explicitly the stationary distribution in a 
matrix product form. We use this algebraic structure to compute physical observables such as the correlation functions and the mean particle current.
\end{abstract}

\vfill\vfill

\newpage
\pagestyle{plain}

\section{Introduction.}

Systems of particles in interaction on a one-dimensional lattice have attracted lots of attention in the last decades. The reason is twofold: 
on one hand they seem to capture the essential physical features of out-of-equilibrium systems, and on the other hand they allow in some particular cases 
for exact computations of physical quantities.
In particular the study of exclusion processes, for which particles experience a hard-core interaction (there is at most one particle on each 
site of the lattice), turned out to be very fruitful. The Asymmetric Simple Exclusion Process (ASEP) is a paradigmatic example of such model
\cite{CMZ,DerrReview}.
The phase diagram of the continuous time model with open boundaries was exactly computed using a matrix product construction of 
the stationary state \cite{DEHP,Sandow94}. The ASEP was also studied in the discrete time setting using various stochastic update rules
\cite{RajewskySSS} (see also for instance \cite{AppertRollandCH} and references therein for more recent developments and applications to traffic flow). 
The matrix product ansatz was successfully used to solve the ASEP with discrete time 
parallel update and sequential update \cite{Hinrichsen,HoneckerP,RajewskySS,RajewskySSS,EvansRS,deGierN,WoelkiS}. 

The reason behind the exact solutions of these models is their integrability: the Markov matrix $M$ ruling their
stochastic dynamics is part of a family of commuting operators generated by a transfer matrix. The transfer matrix is constructed from 
a $R$-matrix satisfying the Yang-Baxter equation \cite{Baxter}.
It is well known that continuous time 
Markov matrices with local stochastic rules can be obtained by taking the first logarithmic derivative of an appropriate transfer matrix.
This mechanism is quite general and relies only on minor assumptions on the $R$-matrix (essentially the regularity property).
It has also been observed that the transfer matrix of the six vertex model can be used to define specific discrete time 
exclusion process \cite{KandelDN,Schutz2,SchutzS}.
However we are lacking a general picture to construct discrete time Markovian dynamics from the transfer matrix, without specific assumptions on 
the $R$-matrix of the model.

We try to address this question in this paper by proposing a generic formalism to obtain Markovian discrete time dynamics from the transfer matrix
of an integrable system. The dynamics can be written easily in terms of local Markovian operators acting on two sites of the lattice.
It can be interpreted as two-step discrete Floquet dynamics, where one pair of sites over two is updated during the first step and the other
pairs during the second step.

We stress that the method can also be used to define integrable discrete time Floquet dynamics for quantum spin chains 
by simply requiring the evolution operator to be unitary (instead of being Markovian) \cite{ProsenVZ}.
We show the efficiency of this general framework by introducing new integrable discrete time generalized exclusion processes, 
where several particles are allowed to occupy the same site instead of a single particle in usual exclusion processes. As far as we know,
these are the first examples of integrable generalized exclusion processes with open boundary conditions.

An advantage of the discrete time picture is that the probabilities $\bra{\cC'}M^n\ket{\cC}$, \textit{i.e} the matrix elements of the propagator of the model
(where $\ket{\cC}$ is the initial configuration, $M$ is the Markov matrix and $\bra{\cC'}$ is the final configuration), can be easily 
interpreted as (equilibrium) partition functions of vertex models in $d=1+1$ dimensions for which a lot of techniques have already been developed.
It should be possible to compute such probabilities in some specific cases for models defined through our formalism. 
   
The outline of the paper is the following. In section \ref{sec:integrable_Floquet} we introduce the transfer matrix formalism and we show how it 
can be used in a general setting to define discrete time Floquet dynamics in the exclusion processes context. This method is valid for both periodic
and open boundary conditions. We explain how the associated stationary state can be computed in a matrix product form. We illustrate the method 
with two examples: the Symmetric Simple Exclusion Process (SSEP) and the Asymmetric Simple Exclusion Process (ASEP).
In section \ref{sec:generalized_exclusion} we use the method to introduce new integrable models with generalized exclusion rules where several 
particles are allowed on the same site. The building blocks of the models, the $R$ and $K$ matrices, are constructed through the fusion procedure.
We present a precise description of the transition probabilities of the Markov matrix. The associated stationary state is exactly computed in 
a matrix product form, by introducing a new technique that we call \textit{fused} matrix ansatz.
We finally end up in section \ref{sec:conclusion} with some concluding remarks and interesting open questions.

\section{Integrable Floquet dynamics from transfer matrix formalism. \label{sec:integrable_Floquet}}

\subsection{Integrability and transfer matrix formalism.}

We give in this subsection a very brief review of integrability in the context of exclusion processes. We introduce the main objects that 
are needed in our construction of discrete time models. The reader is invited to refer to \cite{CRV,VanicatThesis} for details.

\paragraph{Preliminaries.}

We are interested in describing a system of interacting particles on a one dimensional lattice with $L$ sites. Each site of the lattice can 
carry at most $s$ particles. Note that when $s=1$, there is at most one particle per site and this class of model is known as exclusion processes.
Paradigmatic examples of exclusion processes are the ASEP and the SSEP \cite{DerrReview}.
For $s>1$ the exclusion constraint is partially relaxed and the corresponding class of models is called generalized exclusion process.
For each site $i$ of the lattice, we define a local state variable $\tau_i \in \{0,1,\dots,s \}$ denoting
the number of particles lying on site $i$. A configuration of particles on the lattice is thus efficiently encapsulated in a $L$-uplet 
$\bm{\tau}=(\tau_1,\tau_2,\dots,\tau_L)$. We will specify later on a stochastic dynamics on this configuration space, which motivates
the introduction of the quantity $\cP_t(\bm{\tau})$, which stands for the probability to observe the system in the configuration $\bm{\tau}$ at 
time $t$.

The time evolution of $\cP_t(\bm{\tau})$ obeys a master equation. It will be particularly efficient to formulate this master equation in a
matrix form, introducing a Markov matrix and a probability vector. The first step toward this goal is to
associate to each value $\tau=0,\dots,s$ of the local state variable with a local basis vector of $\cV=\CC^{s+1}$
\begin{equation}
\ket{\tau}=(\underbrace{0,\dots,0}_{\tau},1,\underbrace{0,\dots,0}_{s-\tau})^t.
\end{equation}
It allows us to define a probability vector
\begin{equation}
 \ket{\cP_t} = \sum_{\bm{\tau}} \cP_t(\bm{\tau}) \ket{\bm{\tau}},
\end{equation}
where we have constructed a basis vector of $\cV^{\otimes L}$ associated to each configuration $\bm{\tau}$ as tensor product 
of elementary vectors
\begin{equation}
 \ket{\bm{\tau}} = \ket{\tau_1} \otimes \ket{\tau_2} \otimes \cdots \otimes \ket{\tau_L}.
\end{equation}
The time evolution of the probability vector can be encoded in a master equation that takes slightly different forms depending if the stochastic 
model describes \textit{continuous-time} or \textit{discrete-time} dynamics. For \textit{continuous-time} dynamics, the master equation reads
\begin{equation}
 \frac{d\ket{\cP_{t}}}{dt} = M\ket{\cP_t},
\end{equation}
where $M$ is a \textit{continuous-time} Markov matrix (acting on the space $\cV^{\otimes L}$) with non-negative off-diagonal entries
and whose sum of the entries of each column vanishes.

In the case of a \textit{discrete-time} dynamics, the master equation reads
\begin{equation}
 \ket{\cP_{t+1}} = M\ket{\cP_t},
\end{equation}
where $M$ is a \textit{discrete-time} Markov matrix with non-negative entries whose sum over each column is one.
The present paper will be essentially devoted to the construction of integrable \textit{discrete-time} Markov matrices $M$ 
from the transfer matrix approach. We will also be interested in computing exactly in a matrix product form the stationary state\footnote{The Perron-Frobenius theorem states 
that this stationary state exists and is unique if the Markov matrix $M$ is irreducible.}
\footnote{The stationary equation \eqref{eq:steady} is written in the context of discrete-time models. In the continuous-time context it takes
the form $M\steady=0$.} $\steady$, satisfying 
\begin{equation} \label{eq:steady}
 M \steady = \steady. 
\end{equation} 

\paragraph{Periodic boundary conditions.}

The building block of integrable models with periodic boundary conditions is a matrix $R(z)$ depending on a spectral parameter $z$ and acting 
on two components $\cV\otimes\cV$ of the tensor space $\cV$ (\textit{i.e} on two sites of the lattice).

The key equation for integrability in the periodic boundary case is the Yang-Baxter equation
\begin{equation} \label{eq:YBE}
 R_{12}\left(\frac{z_1}{z_2}\right) R_{13}\left(\frac{z_1}{z_3}\right) R_{23}\left(\frac{z_2}{z_3}\right) = 
 R_{23}\left(\frac{z_2}{z_3}\right) R_{13}\left(\frac{z_1}{z_3}\right) R_{12}\left(\frac{z_1}{z_2}\right).
\end{equation}
The previous equation holds in $\cV\otimes\cV\otimes\cV$. The subscripts indicate on which components of the tensor space 
the matrix $R$ is acting non-trivially. For instance $R_{12}(z)=R(z)\otimes 1$, $R_{23}(z)=1 \otimes R(z)$,...
We also require that the matrix $R(z)$ satisfies further properties.
We are in this paper interested in constructing integrable discrete time Markov matrices, \textit{i.e} matrices with non-negative entries whose
sum on each column is equal to one.
We will therefore consider $R$-matrices satisfying the Markovian property
\begin{equation} \label{eq:markovian}
 \bra{\sigma} \otimes \bra{\sigma} R(z) = \bra{\sigma} \otimes \bra{\sigma}
\end{equation}
where $\bra{\sigma} = \sum_{v=0}^{s}\bra{v}$ is the row vector with all entries equal to one.
The latter property is nothing else than imposing that the sum of the entries of each column is equal to one. In our construction the $R$-matrix 
will indeed play the role of a local Markovian operator, see for instance \eqref{eq:definition_U}.
We will also require that $R$-matrices satisfy the regularity property
\begin{equation} \label{eq:regularity}
 R(1)=P
\end{equation}
where $P$ is the permutation operator in $\cV\otimes \cV$, \textit{i.e} we have 
$P \ket{\tau} \otimes \ket{\tau'} = \ket{\tau'} \otimes \ket{\tau}$ for all $\tau,\tau'=0,\dots,s$. In the continuous time setting, 
it is known that this property ensures that one obtains a continuous time Markov matrix (or a quantum Hamiltonian in the context of spin chains) 
acting locally on the lattice from the transfer matrix constructed below. In the discrete time setting, we will show that this property allows us to
simplify the expression of the transfer matrix to end up with a Markov matrix which has an easy interpretation in the parallel update picture.
We also require the $R$-matrix to satisfy the unitarity property 
\begin{equation} \label{eq:unitarity}
 R_{12}(z) R_{21}\left(\frac{1}{z}\right) = 1
\end{equation}
where $R_{21}(z)=PR_{12}(z)P$. This property is used to show the commutation of the transfer matrix below\footnote{From this perspective it 
is not necessary to require unitarity, the $R$-matrix only needs to be invertible.} and it appears as a consistency relation when 
constructing the stationary state in a matrix product form, see subsection \ref{subsec:stationary_state}.

\begin{remark}
 Up to now, we wrote all the definitions and properties about the $R$-matrix assuming that the spectral parameter is \textup{multiplicative}.
 We can easily translate them to an \textup{additive} spectral parameter:
 \begin{itemize}
  \item the Yang-Baxter equation reads
  \begin{equation} \label{eq:YBE_additive}
  R_{12}\left(z_1-z_2\right) R_{13}\left(z_1-z_3\right) R_{23}\left(z_2-z_3\right) = 
 R_{23}\left(z_2-z_3\right) R_{13}\left(z_1-z_3\right) R_{12}\left(z_1-z_2\right)  
  \end{equation}
  \item the Markovian property reads $\bra{\sigma} \otimes \bra{\sigma} R(z) = \bra{\sigma} \otimes \bra{\sigma}$
  \item the regularity property reads $R(0)=P$
  \item the unitarity property reads $R_{12}(z) R_{21}\left(-z\right) = 1$
 \end{itemize}
\end{remark}

We are now equipped to construct the inhomogeneous transfer matrix for a model with periodic boundary conditions
\begin{equation} \label{eq:inhomogeneous_transfer_matrix_periodic}
 t(z|\mathbf{z})=tr_0\left( R_{0L}\left(\frac{z}{z_L}\right) \dots R_{02}\left(\frac{z}{z_2}\right)R_{01}\left(\frac{z}{z_1}\right) \right).
\end{equation}
The transfer matrix acts on the tensor space $\cV^{\otimes L}$ (\textit{i.e} on the whole lattice). The subscripts 
indicate on which components of the tensor space (\textit{i.e} on which sites of the lattice) the operators are acting non-trivially. 
Note that an auxiliary space $0$ is introduced and traced out at the end. 
The main feature of this transfer matrix is that it commutes for different values of the spectral parameter
\begin{equation} \label{eq:transfer_matrix_periodic_commutation}
 [t(x|\mathbf{z}),t(y|\mathbf{z})]=0.
\end{equation}
It can be shown using the Yang-Baxter equation \eqref{eq:YBE} and the unitarity property \eqref{eq:unitarity}.

The transfer matrix is commonly used in the context of exclusion processes (respectively of quantum spin chains) without 
inhomogeneity parameters, \textit{i.e} with all $z_i=1$. In this setting the first logarithmic derivative of the transfer matrix w.r.t the 
spectral parameter generates the continuous time Markov matrix of the exclusion process (or respectively the Hamiltonian of the quantum 
spin chain). The higher order logarithmic derivatives generates other local charges which commute with the Markov matrix (respectively with the 
Hamiltonian) because of the commutation relation \eqref{eq:transfer_matrix_periodic_commutation}.

In the next subsection \ref{subsec:parallel_update}, we will use the inhomogeneous transfer matrix with very particular inhomogeneity parameters 
(different from $1$) to define a discrete time Markov matrix.

\begin{remark}
 In the case of an \textup{additive} spectral parameter, the inhomogeneous transfer matrix reads
 \begin{equation}
 t(z|\mathbf{z})=tr_0\left( R_{0L}\left(z-z_L\right) \dots R_{02}\left(z-z_2\right)R_{01}\left(z-z_1\right) \right).
\end{equation}
\end{remark}

\paragraph{Open boundary conditions.}

In the case of a model with open boundary conditions, the building blocks of integrable models (together with the $R$-matrix already introduced 
in the last paragraph) are the reflection matrices $K(z)$ and $\overline{K}(z)$, both depending on the spectral parameter $z$ and acting on a
single copy $\cV$ of the tensor space (\textit{i.e} on a single site of the lattice). The matrix $K(z)$ is associated to the left boundary and
satisfies the reflection equation
 \begin{equation} \label{eq:reflection_equation}
  R_{12}\left(\frac{z_1}{z_2}\right) K_1(z_1) R_{21}(z_1 z_2) K_2(z_2) =
  K_2(z_2) R_{12}(z_1 z_2) K_1(z_1) R_{21} \left(\frac{z_1}{z_2}\right).
 \end{equation}
The previous equation holds in $\cV\otimes \cV$. Once again the subscripts indicate on which components of the tensor space
the operators are acting non-trivially. For instance $K_1(z) = K(z) \otimes 1$ and $K_2(z) = 1 \otimes K(z)$.
The matrix $\overline{K}(z)$ is associated to the right boundary and satisfies the reversed reflection equation
  \begin{equation} \label{eq:reflection_equation_reversed}
  R_{12}\left(\frac{z_1}{z_2}\right)^{-1} \overline K_1(z_1) R_{21}(z_1 z_2)^{-1} \overline K_2(z_2) =
  \overline K_2(z_2) R_{12}(z_1 z_2)^{-1} \overline K_1(z_1) R_{21}\left(\frac{z_1}{z_2}\right)^{-1}.
 \end{equation}
 
We require also that the $K$-matrices satisfy further properties. We recall that we are interested in constructing discrete time Markov matrices. 
We therefore impose that the $K$-matrices fulfill the Markovian property
\begin{equation} \label{eq:markovian_K}
 \bra{\sigma}K(z)=\bra{\sigma} \quad \mbox{and} \quad \bra{\sigma}\overline{K}(z)=\bra{\sigma}
\end{equation}
which is nothing else than demanding that the sum of the entries on each column of the boundary matrices is equal to one.
In our construction the $K$-matrices will indeed play the role of local Markovian operators on the boundaries.
We also require the boundary matrices to satisfy the regularity property
\begin{equation} \label{eq:regularity_K}
 K(1) = 1 \quad \mbox{and} \quad \overline{K}(1) = 1
\end{equation}
and the unitarity property
\begin{equation} \label{eq:unitarity_K}
 K(z)K\left(\frac{1}{z}\right) = 1 \quad \mbox{and} \quad \overline{K}(z)\overline{K}\left(\frac{1}{z}\right) = 1.
\end{equation}

\begin{remark}
 Once again we wrote all the definitions and properties about the $K$-matrices assuming that the spectral parameter is \textup{multiplicative}.
 We can easily translate them to an \textup{additive} spectral parameter (we give them for the matrix $K(z)$ but similar ones hold also 
 for the matrix $\overline{K}(z)$):
 \begin{itemize}
  \item the reflection equation reads
  \begin{equation} \label{eq:reflection_equation_additive}
R_{12}\left(z_1-z_2\right) K_1(z_1) R_{21}(z_1+z_2) K_2(z_2) =
  K_2(z_2) R_{12}(z_1+z_2) K_1(z_1) R_{21} \left(z_1-z_2\right)  
  \end{equation}
  \item the Markovian property reads $\bra{\sigma} K(z) = \bra{\sigma}$
  \item the regularity property reads $K(0)=1$
  \item the unitarity property reads $K(z)K(-z) = 1$
 \end{itemize}
\end{remark}

We are now in position to define the inhomogeneous transfer matrix for a model with open boundary conditions \cite{sklyanin}
 \begin{equation} \label{eq:inhomogeneous_transfer_matrix_open}
  t(z|\mathbf{z})=tr_0 \left(\widetilde{K}_0(z) R_{0,L}\left(\frac{z}{z_L}\right) \dots R_{0,1}\left(\frac{z}{z_1}\right)
  K_0(z) R_{1,0}(z z_1) \dots R_{L,0}(z z_L) \right),
 \end{equation}
 where the dual boundary matrix $\widetilde{K}(z)$ is 
 defined as\footnote{We recall that $\cdot^{t_i}$ denotes the usual matrix transposition in the $i$-th tensor space component}
 \begin{equation} \label{eq:Ktilde_from_Kb}
  \widetilde K_1(z)= tr_0\left(\overline{K}_0\left(\frac{1}{z}\right)\left(\left(R_{0,1}(z^2)^{t_1}\right)^{-1}\right)^{t_1}P_{0,1}\right),
 \end{equation}
  or equivalently 
 \begin{equation} \label{eq:Kb_from_Ktilde}
  \overline{K}_1(z)= tr_0\left( \widetilde{K}_0\left(\frac{1}{z}\right)R_{01}\left(\frac{1}{z^2}\right)P_{01} \right).
 \end{equation}

 \begin{remark}
  The matrix $\widetilde{K}(z)$ satisfies the dual reflection equation
  \begin{equation} \label{eq:reflection_equation_dual}
   \widetilde{K}_2(z_2)\,\left(R_{21}^{t_1}(z_1z_2)^{-1}\right)^{t_1}\,\widetilde{K}_1(z_1)\,R_{21}\left(\frac{z_2}{z_1}\right)
   =R_{12}\left(\frac{z_2}{z_1}\right)\,\widetilde{K}_1(z_1)\,\left(R_{12}^{t_2}(z_1z_2)^{-1}\right)^{t_2}\,\widetilde{K}_2(z_2).
  \end{equation}
 \end{remark}
 
Thanks to the Yang-Baxter equation \eqref{eq:YBE} and to the reflection equations \eqref{eq:reflection_equation} 
and \eqref{eq:reflection_equation_reversed} (or equivalently \eqref{eq:reflection_equation_dual}), it is possible to show \cite{sklyanin} that
the transfer matrix commutes for different values of the spectral parameter
\begin{equation} \label{eq:transfer_matrix_open_commutation}
 [t(x|\mathbf{z}),t(y|\mathbf{z})]=0.
\end{equation}

In the context of continuous time exclusion processes or of quantum spin chains, the transfer matrix is often used with inhomogeneity parameters 
$z_i=1$. In this case the derivative with respect to the spectral parameter gives a continuous time Markov matrix
(respectively a quantum spin chain Hamiltonian) with two-site local updating rules (respectively with nearest neighbor interactions).
We are going to see in the next subsection \ref{subsec:parallel_update} that the inhomogeneity parameters can be efficiently used to define 
discrete time processes from the transfer matrix. We stress that the construction is quite general in the sense that it does not rely on a 
particular choice for the $R$-matrix and the $K$-matrices (they only have to fulfill the properties mentioned above).

\begin{remark}
 In the \textup{additive} spectral parameter picture, the inhomogeneous transfer matrix reads
 \begin{equation} 
  t(z|\mathbf{z})=tr_0 \left(\widetilde{K}_0(z) R_{0,L}(z-z_L) \dots R_{0,1}(z-z_1)
  K_0(z) R_{1,0}(z+z_1) \dots R_{L,0}(z+z_L) \right)
 \end{equation}
 and the matrices $\overline{K}$ and $\widetilde{K}$ are related for instance through the equation
 \begin{equation}
  \overline{K}_1(z)= tr_0\left( \widetilde{K}_0(-z)R_{01}(-2z)P_{01} \right).
 \end{equation}
\end{remark}

\subsection{Definition of the process. \label{subsec:parallel_update}}

The goal of this subsection is to use the transfer matrix to define a discrete time Markov matrix. The general idea is to specify particular values 
of the spectral parameter $z$ and of the inhomogeneity parameters $z_i$ in the transfer matrix to obtain Markovian dynamics on the lattice with
simple stochastic rules. Several different approaches have already been proposed in the literature.

Among the recent developments we can mention the work \cite{KunibaMMO}
where a large class of integrable discrete time exclusion processes were introduced using particular values for the inhomogeneity parameters 
(this was defined for periodic boundary conditions). It would be interesting to investigate what is the physical interpretation, in terms of 
transition probabilities on the lattice, of all these models. In \cite{CrampeMRV15inhomogeneous} the transfer matrix was used, 
without specifying any inhomogeneity parameters,
to define a discrete time process in the specific case of the totally asymmetric exclusion process for both periodic and open boundary conditions. 
A physical interpretation in terms of a sequential update
was provided. Note that a similar result has been obtained earlier in the periodic homogeneous case in \cite{GM1}. 

In fact the use of the transfer matrix approach in the definition of discrete time exclusion processes follows the pioneering work 
of Baxter \cite{Baxter}. It was soon realized that the inhomogeneous transfer matrix of the six-vertex model can be used 
to define an asymmetric exclusion process with discrete time parallel update on the periodic lattice \cite{KandelDN,SchutzS,Schutz2,Schutz}.
Our construction is inspired by these works and aims to provide a general framework (independent of the $R$ matrix considered) to define
an integrable discrete time Markov matrix from the transfer matrix machinery of integrable systems. Moreover we generalize it also for 
open boundary conditions using the Sklyanin transfer matrix \cite{sklyanin}. The discrete time dynamics that we obtain for both periodic and 
open boundary conditions appears to be two-step Floquet dynamics. Note that another approach has been developed recently in \cite{FloquetInt}
to construct integrable continuous time Floquet dynamics for quantum systems from the transfer matrix formalism.

\paragraph{Periodic boundary conditions.}

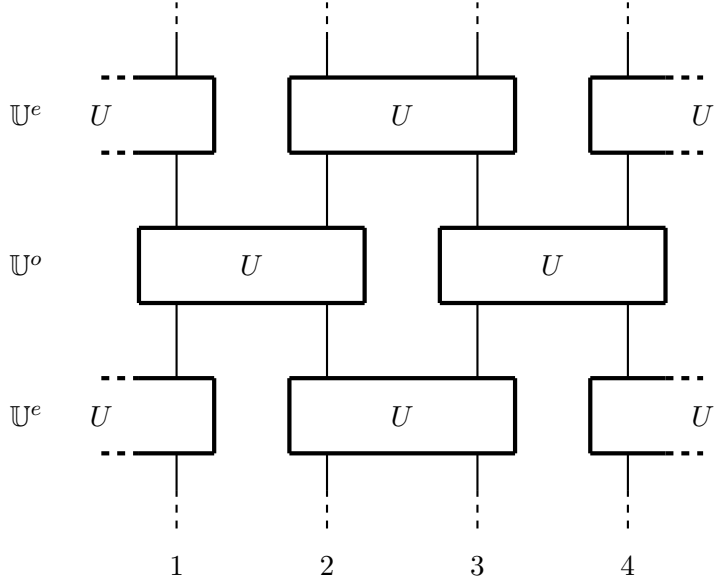
\begin{figure}[htb]
\begin{center}
 \begin{tikzpicture}[scale=1.0]
\foreach \i in {0,2,...,6}
{\draw[thick,dashed] (\i,0) -- (\i,0.5) ;
\draw[thick] (\i,0.5) -- (\i,1) ;
\draw[thick] (\i,2) -- (\i,3) ;
\draw[thick] (\i,4) -- (\i,5) ;
\draw[thick] (\i,6) -- (\i,6.5) ;
\draw[thick,dashed] (\i,6.5) -- (\i,7) ;}
\foreach \i in {1,2,5,6}
{\draw[ultra thick,dashed] (-1,\i) -- (-0.5,\i) ;
\draw[ultra thick] (-0.5,\i) -- (0.5,\i) ;
\draw[ultra thick] (1.5,\i) -- (4.5,\i) ;
\draw[ultra thick] (5.5,\i) -- (6.5,\i) ;
\draw[ultra thick,dashed] (6.5,\i) -- (7,\i) ;}
\foreach \i in {3,4}
{\draw[ultra thick] (-0.5,\i) -- (2.5,\i) ;
\draw[ultra thick] (3.5,\i) -- (6.5,\i) ;}
\foreach \i in {0.5,1.5,4.5,5.5}
{\draw[ultra thick] (\i,1) -- (\i,2) ;
\draw[ultra thick] (\i,5) -- (\i,6) ;}
\foreach \i in {-0.5,2.5,3.5,6.5}
{\draw[ultra thick] (\i,3) -- (\i,4) ;}
\node at (0,-0.5) [] {$1$}; \node at (2,-0.5) [] {$2$}; \node at (4,-0.5) [] {$3$}; \node at (6,-0.5) [] {$4$}; 
\node at (-1,1.5) [] {$U$}; \node at (-1,5.5) [] {$U$};
\node at (3,1.5) [] {$U$}; \node at (3,5.5) [] {$U$}; \node at (7,1.5) [] {$U$}; \node at (7,5.5) [] {$U$};
\node at (1,3.5) [] {$U$}; \node at (5,3.5) [] {$U$}; 
\node at (-2,1.5) [] {$\UU^{e}$}; \node at (-2,5.5) [] {$\UU^{e}$};
\node at (-2,3.5) [] {$\UU^{o}$};
 \end{tikzpicture}
 \end{center}
 \caption{Pictorial representation of the discrete time dynamics for periodic boundary conditions. Note that the time flows upward.}
 \label{fig:periodic}
\end{figure}

To define our process on the periodic lattice, we need to consider a lattice with an even number of sites $L$. 
We fix $z_1=z_3=\dots=z_{L-1}=\frac{1}{\kappa}$ and $z_2=z_4=\dots=z_L=\kappa$. The inhomogeneous transfer matrix then takes
the following staggered form
\begin{equation} \label{eq:transfer_matrix_periodic}
 t(z)=tr_0\left( R_{0L}\left(\frac{z}{\kappa}\right)R_{0L-1}\left(z\kappa\right) \dots R_{02}\left(\frac{z}{\kappa}\right)R_{01}\left(z\kappa\right) \right).
\end{equation}
Note that a similar staggered construction already appeared in a monodromy matrix in \cite{FaddeevV}.
Let us introduce the local Markovian operator (acting on two sites) 
\begin{equation} \label{eq:definition_U}
 U = \check R(\kappa^2)
\end{equation}
where $\check R(z)=P.R(z)$. Note that $U$ also fulfill the Markovian property $\bra{\sigma}U=\bra{\sigma}$.
A straightforward computation\footnote{We use essentially the regularity and the unitarity properties of the $R$-matrix.} yields 
\begin{equation}
 t\left(\frac{1}{\kappa}\right)= P_{1L}P_{1L-1}\dots P_{13}P_{12} \times U_{L-1,L}^{-1} U_{L-3,L-2}^{-1} \dots U_{1,2}^{-1}
\end{equation}
and 
\begin{equation}
 t(\kappa) = P_{1L}P_{1L-1}\dots P_{13}P_{12} \times U_{L-2,L-1}U_{L-4,L-3} \dots U_{2,3}U_{L,1}.
\end{equation}
We can thus define a Markov matrix $M$ in the following way
\begin{equation} \label{eq:Markov_matrix_periodic}
 M=t\left(\frac{1}{\kappa}\right)^{-1}t(\kappa) = \UU^{o} \UU^{e}
\end{equation}
with\footnote{The indices in the formula have to be understood modulo $L$, for instance $0\equiv L$ and $L+1\equiv 1$.}
\begin{equation} \label{eq:Floquet_operators_periodic}
 \UU^{o} = \prod_{k=1}^{L/2} U_{2k-1,2k} \quad \mbox{and} \quad \UU^{e} = \prod_{k=1}^{L/2} U_{2k,2k+1}.
\end{equation}
The property that expression \eqref{eq:Markov_matrix_periodic} together with \eqref{eq:Floquet_operators_periodic} fulfills the requirement of 
a Markov matrix, \textit{i.e} non-negativity and sum to one property of the entries, follows from the fact that the entries of the local 
operator $U$ are non-negative and sum to one on each column (thanks to the Markovian property \eqref{eq:markovian} of the R-matrix).
Note that the local operators $U$ involved in the definition of $\UU^{o}$ all commute one with each other and thus the order in the product is 
not important. The same is true for $\UU^{e}$.
From the expression \eqref{eq:Markov_matrix_periodic} it is clear that the Markov matrix $M$ can be interpreted as describing a Floquet 
dynamics composed of two steps: the first one embodied in $\UU^{e}$ and the other one in $\UU^{o}$. We stress that 
$\UU^{e}$ and $\UU^{o}$ do not commute. The operator $\UU^{e}$ updates every pair of consecutive sites with the first site located at an even position.
The local stochastic update is encoded in the matrix $U$ acting on $\cV\otimes\cV$. We will give precise examples of such
matrices in subsections \ref{subsec:examples} and \ref{subsec:fusion}.
The operator $\UU^{o}$ updates every pair of consecutive sites with the first site located at an odd position. A pictorial representation of 
the discrete time dynamics is given in figure \ref{fig:periodic}.
In the exclusion process literature this kind of dynamics is sometimes referred to as discrete time parallel update, 
see for instance \cite{Schutz2, Schutz}.
The same structure also appeared in the propagator of the quantum Hirota model \cite{FaddeevV} 
and in a deterministic cellular automaton \cite{MedenjakKP}.

The model is said to be integrable because the Markov matrix $M$ commutes with a whole family of operators generated by $t(z)$
\begin{equation}
 [M,t(z)] = 0
\end{equation}
which is direct consequence of the commutation relation \eqref{eq:transfer_matrix_periodic_commutation} 
and of the definition \eqref{eq:Markov_matrix_periodic}.

\paragraph{Open boundary conditions.}

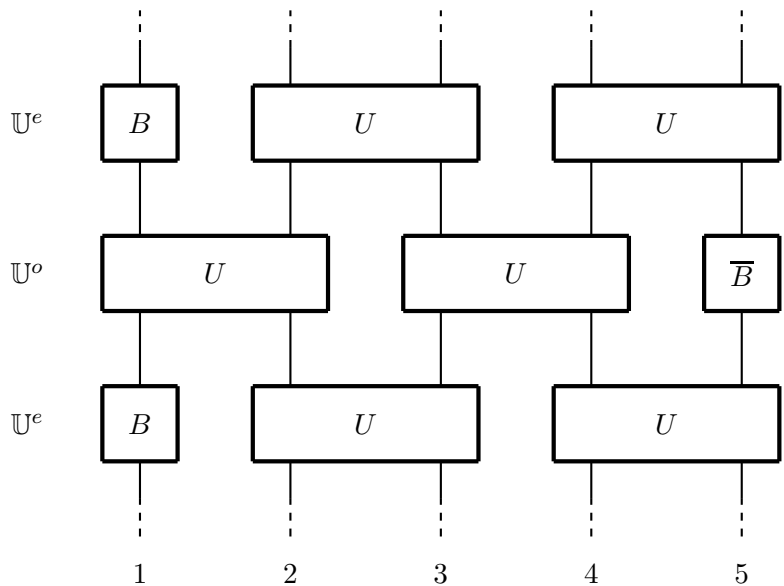
\begin{figure}[htb]
\begin{center}
 \begin{tikzpicture}[scale=1.0]
\foreach \i in {0,2,...,8}
{\draw[thick,dashed] (\i,0) -- (\i,0.5) ;
\draw[thick] (\i,0.5) -- (\i,1) ;
\draw[thick] (\i,2) -- (\i,3) ;
\draw[thick] (\i,4) -- (\i,5) ;
\draw[thick] (\i,6) -- (\i,6.5) ;
\draw[thick,dashed] (\i,6.5) -- (\i,7) ;}
\foreach \i in {1,2,5,6}
{\draw[ultra thick] (-0.5,\i) -- (0.5,\i) ;
\draw[ultra thick] (1.5,\i) -- (4.5,\i) ;
\draw[ultra thick] (5.5,\i) -- (8.5,\i) ;}
\foreach \i in {3,4}
{\draw[ultra thick] (-0.5,\i) -- (2.5,\i) ;
\draw[ultra thick] (3.5,\i) -- (6.5,\i) ;
\draw[ultra thick] (7.5,\i) -- (8.5,\i) ;}
\foreach \i in {-0.5,0.5,1.5,4.5,5.5,8.5}
{\draw[ultra thick] (\i,1) -- (\i,2) ;
\draw[ultra thick] (\i,5) -- (\i,6) ;}
\foreach \i in {-0.5,2.5,3.5,6.5,7.5,8.5}
{\draw[ultra thick] (\i,3) -- (\i,4) ;}
\node at (0,-0.5) [] {$1$}; \node at (2,-0.5) [] {$2$}; \node at (4,-0.5) [] {$3$}; \node at (6,-0.5) [] {$4$}; \node at (8,-0.5) [] {$5$};
\node at (0,1.5) [] {$B$}; \node at (0,5.5) [] {$B$};
\node at (3,1.5) [] {$U$}; \node at (3,5.5) [] {$U$}; \node at (7,1.5) [] {$U$}; \node at (7,5.5) [] {$U$};
\node at (1,3.5) [] {$U$}; \node at (5,3.5) [] {$U$};
\node at (8,3.5) [] {$\overline{B}$}; 
\node at (-1.5,1.5) [] {$\UU^{e}$}; \node at (-1.5,5.5) [] {$\UU^{e}$};
\node at (-1.5,3.5) [] {$\UU^{o}$};
 \end{tikzpicture}
 \end{center}
 \caption{Pictorial representation of the discrete time dynamics for open boundary conditions. Note that the time flows upward.}
 \label{fig:open}
\end{figure}

To define our process on a lattice with open boundaries, we need to consider a lattice with an odd number of sites $L$. 
The idea is essentially the same as for the periodic boundary conditions case.
We fix $z_1=z_3=\dots=z_L=\kappa$ and $z_2=z_4=\dots=z_{L-1}=\frac{1}{\kappa}$. The inhomogeneous transfer matrix then takes
the following staggered form
\begin{equation} \label{eq:transfer_matrix_open}
  t(z)=tr_0 \left(\widetilde{K}_0(z) R_{0L}\left(\frac{z}{\kappa}\right)\dots R_{02}\left(z\kappa\right) R_{01}\left(\frac{z}{\kappa}\right)
  K_0(z) R_{10}(z\kappa)R_{20}\left(\frac{z}{\kappa}\right) \dots R_{L0}(z\kappa) \right).
 \end{equation}
A direct computation yields
\begin{equation}
 t(\kappa) = B_1 U_{23}U_{45}\dots U_{L-1,L}\, U_{12}U_{34} \dots U_{L-2,L-1} \overline{B}_L,
\end{equation}
where $B=K(\kappa)$ and $\overline{B}=\overline{K}(1/\kappa)$. Once again we can define a Markov matrix as
\begin{equation} \label{eq:Markov_matrix_open}
 M = t(\kappa) = \UU^{e} \UU^{o}
\end{equation}
with operators $\UU^{e}$ and $\UU^{o}$ having similar expression as in the periodic case but with additional boundary terms
\begin{equation}
 \UU^{o} = \prod_{k=1}^{\frac{L-1}{2}} U_{2k-1,2k} \, \overline{B}_L \quad \mbox{and} \quad \UU^{e} =  B_1 \prod_{k=1}^{\frac{L-1}{2}} U_{2k,2k+1}.
\end{equation}
The local operators involved in $\UU^{o}$ (respectively in $\UU^{e}$) commute one with each other because they are acting on different sites of 
the lattice. $\UU^{o}$ realizes a stochastic update on 
every pair of consecutive sites with the first site located at an odd position, and also a stochastic update on the last site $L$ (describing
an interaction with reservoir at the right boundary). $\UU^{e}$ similarly realizes a stochastic update on 
every pair of consecutive sites with the first site located at an even position, and also a stochastic update on the first site $1$ (describing
an interaction with reservoir at the left boundary). Note that the operators $\UU^{o}$ and $\UU^{e}$ do not commute and can be interpreted as 
describing a two-step Floquet dynamics. A pictorial representation of the discrete time dynamics is given in figure \ref{fig:open}.
This discrete time update is very similar to the \textit{sublattice-parallel update} \cite{RajewskySSS}: the difference is that in the present model
the size of the lattice $L$ has to be \textit{odd} instead of \textit{even} in the sublattice-parallel update with open boundaries.

Once again the model is said to be integrable because the Markov matrix $M$ commutes with a whole family of operators generated by the transfer 
matrix
\begin{equation}
 [M,t(z)]=0.
\end{equation}
The goal is now to use the integrable machinery to handle exact computations for these models.

\subsection{Stationary distribution in a matrix product form. \label{subsec:stationary_state}}

It turns out that the integrability framework can be efficiently exploited to construct exactly in a matrix product form the stationary state 
of the models. The general method has been proposed in \cite{Sasamoto2,CRV} and we sketch here the main points.
We will focus on the open boundaries case but the periodic boundaries case can be treated similarly and we just provide the solution at 
the end of the subsection.
We introduce a $(s+1)$-component 
vector $\mathbf{A}(z)$ depending on a spectral parameter and with algebraic-valued (\textit{i.e} non-commutative) entries. These algebraic 
elements will be basically the matrices entering the matrix product construction of the stationary state.

The construction relies essentially on two key relations. The Zamolodchilov-Faddeev (ZF) relation encodes the commutation relations of the 
matrix ansatz algebra
 \begin{equation}\label{eq:ZF}
 \check{R}\left(\frac{z_1}{z_2}\right) \mathbf{A}(z_1) \otimes \mathbf{A}(z_2) = \mathbf{A}(z_2) \otimes \mathbf{A}(z_1).
 \end{equation}
 The associativity of the algebra is ensured by the Yang-Baxter equation \eqref{eq:YBE}. Another consistency relation is ensured by the 
 unitarity property \eqref{eq:unitarity}.
 
 The Ghoshal-Zamolodchilov (GZ) relations encode the action of the matrix ansatz algebra on the boundary vectors $\llangle W|$ and $|V\rrangle$ (which are 
 used to contract the matrix product into a scalar)
  \begin{equation} \label{eq:GZ}
  \llangle W| \left( K(z)\mathbf{A}\left(\frac{1}{z}\right)-\mathbf{A}(z) \right)=0, \qquad 
  \left(\overline{K}(z)\mathbf{A}\left(\frac{1}{z}\right)-\mathbf{A}(z) \right)|V\rrangle = 0.
 \end{equation}
 The consistency between these actions and the commutation relations of the matrix ansatz algebra (\textit{i.e} the ZF relation) is ensured 
 by the reflection equations \eqref{eq:reflection_equation} and \eqref{eq:reflection_equation_reversed}.
 
 We can now introduce the matrix product state
 \begin{equation} \label{eq:inhomogeneous_ground_state}
 \ket{\cS(z_1,z_2,\dots,z_L)}=\frac{1}{Z_L(z_1,z_2,\dots,z_L)}
 \llangle W| \mathbf{A}(z_1) \otimes \mathbf{A}(z_2) \otimes \dots \otimes \mathbf{A}(z_L) |V\rrangle,
\end{equation}
with $Z_L(z_1,z_2,\dots,z_L) = \llangle W|\bm{C}(z_1)\bm{C}(z_2) \cdots \bm{C}(z_L)|V\rrangle$ where the algebraic element $\bm{C}(z)$ is defined
as the sum of the components of the vector $\bm{A}(z)$
\begin{equation}
 \bm{C}(z)=\bra{\sigma}\bm{A}(z).
\end{equation}
It has been shown in \cite{CRV} that if the ZF and GZ relations \eqref{eq:ZF} and \eqref{eq:GZ} are fulfilled, then $\ket{\cS(z_1,z_2,\dots,z_L)}$ 
satisfies
\begin{equation}
t(z_i)\ket{\cS(z_1,z_2,\dots,z_L)} = \ket{\cS(z_1,z_2,\dots,z_L)}, \quad \mbox{for} \quad 1\leq i \leq L,
\end{equation}
where $t(z)$ is the inhomogeneous transfer matrix defined in \eqref{eq:inhomogeneous_transfer_matrix_open} with inhomogeneity parameters
$z_1,\dots,z_L$. On specific models, it is even possible to prove using symmetry and degree arguments that $\ket{\cS(z_1,z_2,\dots,z_L)}$ is 
an eigenvector of $t(z)$ for all $z$ (but with an eigenvalue possibly different from $1$), see for instance \cite{CRV,VanicatThesis}.

For our process we recall that we took a specific staggered picture for the inhomogeneity parameters, see subsection \ref{subsec:parallel_update}.
The steady state of the process is thus given in the staggered form (similar to the transfer matrix \eqref{eq:transfer_matrix_open})
\begin{equation}
 \steady = \frac{1}{Z_L} \llangle W|\mathbf{A}(\kappa)\otimes\mathbf{A}\left(\frac{1}{\kappa}\right)\otimes\dots\otimes\mathbf{A}(\kappa)|V\rrangle
\end{equation}
where $Z_L= Z_L(\kappa,1/\kappa,\dots,\kappa)$. Let us briefly show that it provides the correct stationary state.
The following relations indeed hold (they are a direct consequence of \eqref{eq:ZF} and \eqref{eq:GZ})
\begin{equation}
 U \mathbf{A}(\kappa)\otimes\mathbf{A}\left(\frac{1}{\kappa}\right)= \mathbf{A}\left(\frac{1}{\kappa}\right) \otimes \mathbf{A}(\kappa), \quad 
 B \bbra{W} \mathbf{A}\left(\frac{1}{\kappa}\right)= \bbra{W} \mathbf{A}(\kappa), \quad 
 \overline{B} \mathbf{A}(\kappa) \kket{V} =  \mathbf{A}\left(\frac{1}{\kappa}\right)\kket{V}.
\end{equation}
Using these properties, it is then easy to check that
\begin{equation}
 \UU^{o} \steady = \ket{\cS'} \quad \mbox{and} \quad \UU^{e} \ket{\cS'} = \steady, 
\end{equation}
where
\begin{equation} \label{eq:stationary_state_bis}
 \ket{\cS'} = \frac{1}{Z_L} \llangle W|\mathbf{A}\left(\frac{1}{\kappa}\right)\otimes \mathbf{A}(\kappa)\otimes\dots\otimes\mathbf{A}\left(\frac{1}{\kappa}\right)|V\rrangle.
\end{equation}
It thus implies $M\steady = \steady$. Note that this kind of staggering mechanism for the stationary state has already appeared in \cite{ProsenB}. 

\begin{remark}
 For an \textup{additive} spectral parameter, the stationary state is given by 
 \begin{equation}
 \steady = \frac{1}{Z_L} \llangle W|\mathbf{A}(\kappa)\otimes\mathbf{A}(-\kappa)\otimes\dots\otimes\mathbf{A}(\kappa)|V\rrangle
\end{equation}
where $Z_L= Z_L(\kappa,-\kappa,\dots,\kappa)$.
\end{remark}

We recall also that the construction of the stationary state exposed in this subsection holds only for models with open boundary conditions.
For models with periodic boundary conditions, we would need to adapt the construction of the stationary state
\begin{equation}
 \steady_N = \frac{1}{Z_{L,N}} \pi_N\Big[
 tr \left(\mathbf{A}(\kappa)\otimes\mathbf{A}\left(\frac{1}{\kappa}\right)\otimes\dots\otimes \mathbf{A}\left(\frac{1}{\kappa}\right)\right)\Big],
\end{equation}
where $tr(\cdot)$ denotes a trace operator (\textit{i.e} satisfying the cyclic property) on the auxiliary space of the matrix ansatz algebra. 
$\pi_N$ is the projector (in the physical space of configurations) on the sector with $N$ particles on the lattice.
Indeed with periodic boundary conditions, the number 
of particles in the system is often conserved and the configuration space can be decomposed into different sectors (labeled by the number of particles)
that are stable under the action of the Markov matrix.
$Z_{L,N}$ is the appropriate normalization (so that the sum of the components of $\steady_N$ is $1$). 

To ease the presentation, and to avoid dealing with these different sectors, we will focus on systems with open boundary conditions in the 
rest of the paper. We will provide explicit examples of the construction of discrete time models and of the associated stationary state.

\subsection{Examples: the SSEP and ASEP case. \label{subsec:examples}}

In this subsection we apply the generic construction introduced above in subsection \ref{subsec:parallel_update} to the $R$ and $K$ matrices 
associated with the SSEP and the ASEP.
We define the discrete time stochastic dynamics and provide explicitly the transition probabilities between the configurations. The stationary 
states are then computed following the general procedure presented in \ref{subsec:stationary_state}. The matrix ansatz algebra is found to be 
closely related to the one used to solve the continuous time models.

\paragraph{The SSEP case.}

We are first interested in the SSEP, which is a stochastic model where particles can jump to the left or to the right with equal probability
(\textit{i.e} there is no driving force in the bulk) and with an exclusion constraint. 
We introduce an $R$-matrix which reads
in the basis $\ket{0}\otimes\ket{0}$, $\ket{0}\otimes\ket{1}$, $\ket{1}\otimes\ket{0}$ and $\ket{1}\otimes\ket{1}$ (ordered this way)
\begin{equation} \label{eq:SSEP_R}
  R(z) = \begin{pmatrix}
          1 & 0 & 0 & 0 \\ 
          0 & \frac{z}{z+1} & \frac{1}{z+1} & 0 \\
          0 & \frac{1}{z+1} & \frac{z}{z+1} & 0 \\
          0 & 0 & 0 & 1
         \end{pmatrix}.
 \end{equation}
 It satisfies the Yang-Baxter equation \eqref{eq:YBE_additive}, the Markovian, regularity and unitarity properties. The $R$-matrix will be 
 used below to define a local Markovian operator in the bulk of the system. The jumping probabilities to the left and to the right will be equal 
 because of the following property of the $R$-matrix: $PR(z)P=R(z)$.
We also introduce two reflection matrices which read in the basis $\ket{0}$, $\ket{1}$ (ordered this way)
\begin{equation} \label{eq:SSEP_K}
  K(z) = \begin{pmatrix}
          \frac{(c-a)z+1}{(a+c)z+1} & \frac{2cz}{(a+c)z+1} \\
          \frac{2az}{(a+c)z+1} & \frac{(a-c)z+1}{(a+c)z+1}
         \end{pmatrix} \quad \mbox{and} \quad 
  \overline{K}(z) = \begin{pmatrix}
          \frac{(b-d)z-1}{(b+d)z-1} & \frac{2bz}{(b+d)z-1} \\
          \frac{2dz}{(b+d)z-1} & \frac{(d-b)z-1}{(b+d)z-1}
         \end{pmatrix}.
\end{equation}
They satisfy the reflection equation \eqref{eq:reflection_equation_additive} and the reversed reflection equation \eqref{eq:reflection_equation_reversed}
respectively (with additive spectral parameters). They also fulfill the Markovian, regularity and unitarity properties. They will be used to define
local Markovian operators acting on the first and last sites, which describe the interaction with particle reservoirs at the boundaries.
 
Following the general procedure introduced in subsection \ref{subsec:parallel_update} we define the local Markovian operators using the 
$R$ and $K$ matrices
\begin{equation}
 U = \check R(2\kappa), \qquad B=K(\kappa), \qquad \overline{B}=\overline{K}(-\kappa). 
\end{equation}
It allows us to define an integrable discrete time Markov matrix defined by \eqref{eq:Markov_matrix_open}. We now give an explicit description of the 
local stochastic rules of the model.
In the bulk the dynamics encoded by the matrix $U$ explicitly reads
\begin{equation}
 \begin{aligned}
  & 01 \longrightarrow 10 \quad \frac{2\kappa}{2\kappa+1}; \qquad 01 \longrightarrow 01 \quad \frac{1}{2\kappa+1}; \\
  & 10 \longrightarrow 01 \quad \frac{2\kappa}{2\kappa+1}; \qquad 10 \longrightarrow 10 \quad \frac{1}{2\kappa+1}; \\
 \end{aligned}
\end{equation}
Note that the dynamics are symmetric as announced: the particles jump to the left or to the right with the same probability.
On the left boundary the dynamics encoded by the matrix $B$ reads
\begin{equation}
 \begin{aligned}
  & 0 \longrightarrow 1 \quad \frac{2a\kappa}{(a+c)\kappa+1}; \qquad 0 \longrightarrow 0 \quad \frac{(c-a)\kappa+1}{(a+c)\kappa+1}; \\
  & 1 \longrightarrow 0 \quad \frac{2c\kappa}{(a+c)\kappa+1}; \qquad 1 \longrightarrow 1 \quad \frac{(a-c)\kappa+1}{(a+c)\kappa+1}; \\
 \end{aligned}
\end{equation}
It may for instance describe the interaction with a particle reservoir at fixed density.
On the right boundary the dynamics encoded by the matrix $\overline{B}$ reads
\begin{equation}
 \begin{aligned}
  & 0 \longrightarrow 1 \quad \frac{2d\kappa}{(b+d)\kappa+1}; \qquad 0 \longrightarrow 0 \quad \frac{(b-d)\kappa+1}{(b+d)\kappa+1}; \\
  & 1 \longrightarrow 0 \quad \frac{2b\kappa}{(b+d)\kappa+1}; \qquad 1 \longrightarrow 1 \quad \frac{(d-b)\kappa+1}{(b+d)\kappa+1}; \\
 \end{aligned}
\end{equation}

Now that the discrete time model is precisely defined, we can look for its stationary distribution in a matrix product form following 
the general method presented in subsection \ref{subsec:stationary_state}. We only need to construct a vector $\bm{A}(z)$ satisfying the 
ZF \eqref{eq:ZF} and GZ \eqref{eq:GZ} relations (with additive spectral parameters).

For such a purpose, we introduce the vector $\bm{A}(z)$ with algebraic entries 
\begin{equation} \label{eq:SSEP_A}
 \bm{A}(z) = \begin{pmatrix}
              -z+\bm{E} \\ z+\bm{D}
             \end{pmatrix}.
\end{equation}
The matrices $\bm{E}$ and $\bm{D}$ satisfy the following commutation relation
\begin{equation} \label{eq:SSEP_commutation_relation}
\bm{DE}-\bm{ED}=\bm{D}+\bm{E}
\end{equation}
and the relations on the boundary vectors $\bbra{W}$ and $\kket{V}$
\begin{equation} \label{eq:SSEP_relations_boundaries}
\bbra{W}\Big(a\bm{E}-c\bm{D}-1\Big)=0, \qquad \mbox{and} \qquad \Big(b\bm{D}-d\bm{E}-1\Big)\kket{V}=0.
\end{equation}
Note that these algebraic relations are exactly the ones appearing when constructing the stationary state of the continuous time SSEP. It is known that
there exists an explicit representation of $\bm{E}$, $\bm{D}$, $\bbra{W}$ and $\kket{V}$, satisfying \eqref{eq:SSEP_commutation_relation} and 
\eqref{eq:SSEP_relations_boundaries}, as infinite dimensional matrices and vectors.
A direct computation shows that if the relations \eqref{eq:SSEP_commutation_relation} and \eqref{eq:SSEP_relations_boundaries} are fulfilled, then the 
vector $\bm{A}(z)$ satisfies the ZF \eqref{eq:ZF} and GZ \eqref{eq:GZ} relations (together with $\bbra{W}$ and $\kket{V}$) 
with additive spectral parameters.

We stress that, despite the similitude of the underlying algebraic relations, 
the stationary distribution of the present discrete time SSEP model is not the same as the one of the continuous time model because of 
the staggered construction\footnote{The stationary state of the continuous time SSEP would be rather given by
$\steady = \frac{1}{Z_L}\bbra{W} \bm{A}(0) \otimes \cdots \otimes \bm{A}(0) \kket{V}$}
\begin{equation}
 \steady = \frac{1}{Z_L}\bbra{W} \bm{A}(\kappa) \otimes \bm{A}(-\kappa) \otimes \bm{A}(\kappa) \otimes \cdots \otimes \bm{A}(\kappa) \kket{V}.
\end{equation}
We can use this matrix product solution to compute physical quantities. The first step is to compute the normalization $Z_L$. We recall that it is 
defined as
\begin{equation}
 Z_L= Z_L(\kappa,-\kappa,\dots,\kappa) = \bbra{W} \bm{C}(\kappa)\bm{C}(-\kappa)\cdots \bm{C}(\kappa)\kket{V} = \bbra{W}(\bm{E}+\bm{D})^L \kket{V}
\end{equation}
because $\bm{C}(z)=\bra{\sigma}\bm{A}(z)=\bm{E}+\bm{D}$ is independent of $z$.
It is exactly the expression of the normalization for the continuous time SSEP \cite{DerrReview,MartinRev}. We thus have
\begin{equation} \label{eq:SSEP_normalisation}
 Z_L = \frac{(a+c)^L(b+d)^L}{(ab-cd)^L} \frac{\Gamma\left(L+\frac{1}{a+c}+\frac{1}{b+d}\right)}{\Gamma\left(\frac{1}{a+c}+\frac{1}{b+d}\right)}
\end{equation}
where the function Gamma satisfies $\Gamma(z+1)=z\Gamma(z)$. We can also compute the mean particle density on site $i$. 
We have to be careful because of the two-step 
Floquet dynamics: in the stationary regime the system is for one out of two time steps described by the probability vector $\steady$ and for one 
out of two time steps described by the probability vector $\ket{\cS'}$ (defined in \eqref{eq:stationary_state_bis}). In the stationary regime the 
mean particle density (averaged also over the two time steps of the Floquet dynamics) is thus given by
\begin{eqnarray} \label{eq:SSEP_density}
 \langle \tau_i \rangle & = & \frac{1}{2}\frac{1}{Z_L}\bbra{W}(\bm{E}+\bm{D})^{i-1}\big((-1)^{i+1}\kappa+\bm{D}\big)(\bm{E}+\bm{D})^{L-i}\kket{V} \\
 & & + \frac{1}{2}\frac{1}{Z_L}\bbra{W}(\bm{E}+\bm{D})^{i-1}\big((-1)^{i}\kappa+\bm{D}\big)(\bm{E}+\bm{D})^{L-i}\kket{V} \\
 & = & \frac{1}{Z_L}\bbra{W}(\bm{E}+\bm{D})^{i-1}\bm{D}(\bm{E}+\bm{D})^{L-i}\kket{V} \\
 & = & \frac{\frac{a}{a+c}\left(L+\frac{1}{b+d}-i\right)+\frac{d}{b+d}\left(i-1+\frac{1}{a+c}\right)}{L+\frac{1}{a+c}+\frac{1}{b+d}-1},
\end{eqnarray}
where the last equality is obtained using again the results derived for the continuous time SSEP matrix ansatz algebra \cite{DerrReview,MartinRev}. 
It appears to be identical to the continuous time SSEP density. Let us mention here that the similarity of the physical observables 
between the discrete time and continuous time SSEP is specific to this model. We will see in the next paragraph that the results are different 
between discrete and continuous time in the ASEP case. Note that the result \eqref{eq:SSEP_density} can be obtained without using the matrix product
structure because the mean densities satisfy a set of closed equations, similarly to the continuous time case. The matrix product solution provides
nevertheless an efficient tool to compute multi-point correlation functions.

The computation of the mean particle current can also be handled. We have to be careful again because of the Floquet two 
steps dynamics: particles can only jump from site $i$ to site $i+1$, where $i$ is odd (respectively even), during the first time step (respectively
the second time step) encoded by $\UU^{o}$ (respectively by $\UU^{e}$). It implies that when $i$ is odd we have to investigate the action of 
$\UU^{o}$ on sites $i$ and $i+1$ of the vector $\steady$, whereas when $i$ is even we have to investigate the action of $\UU^{e}$ 
on sites $i$ and $i+1$ of the vector $\ket{\cS'}$. It turns out that both cases reduce consistently 
to the same following computation
\begin{eqnarray}
 \langle J \rangle & = & \frac{1}{Z_L}\bbra{W}(\bm{E}+\bm{D})^{i-1}
 \frac{2\kappa}{2\kappa+1}\Big[\big(\kappa+\bm{D}\big)\big(\kappa+\bm{E}\big)-\big(-\kappa+\bm{E}\big)\big(-\kappa+\bm{D}\big)\Big](\bm{E}+\bm{D})^{L-i-1}\kket{V}\nonumber \\
 & = & 2\kappa\frac{Z_{L-1}}{Z_L} \nonumber \\
 & = & 2\kappa\frac{\frac{a}{a+c}-\frac{d}{b+d}}{L+\frac{1}{a+c}+\frac{1}{b+d}-1}.
\end{eqnarray}
Due to the factor $2\kappa$ it differs slightly from the corresponding result for the continuous time SSEP.

\paragraph{The ASEP case.}

We are now interested in the ASEP, which is a stochastic model where particles can jump to the left or to the right with an asymmetric probability
(which can describe a driving force in the bulk) and with an exclusion constraint. 
We introduce a $R$-matrix which reads
in the basis $\ket{0}\otimes\ket{0}$, $\ket{0}\otimes\ket{1}$, $\ket{1}\otimes\ket{0}$ and $\ket{1}\otimes\ket{1}$ (ordered this way)
\begin{equation} \label{eq:ASEP_R}
  R(z) = \begin{pmatrix}
          1 & 0 & 0 & 0 \\ 
          0 & \frac{(1-z)t^2}{1-t^2z} & \frac{z(1-t^2)}{1-t^2z} & 0 \\
          0 & \frac{1-t^2}{1-t^2z} & \frac{(1-z)}{1-t^2z} & 0 \\
          0 & 0 & 0 & 1
         \end{pmatrix}
 \end{equation}
It satisfies the Yang-Baxter equation \eqref{eq:YBE}, the Markovian \eqref{eq:markovian}, regularity \eqref{eq:regularity}
and unitarity \eqref{eq:unitarity} properties. The $R$-matrix will be again 
 used below to define a local Markovian operator in the bulk of the system. 
 Note that the R-matrix of the SSEP \eqref{eq:SSEP_R} can be recovered from the R-matrix of the ASEP \eqref{eq:ASEP_R} taking the limit 
 $R^{SSEP}(z)=\lim_{h\rightarrow 0} R^{ASEP}(e^{hz})|_{t^2=e^h}$.
 
We also introduce two reflection matrices which read in the basis $\ket{0}$, $\ket{1}$ (ordered this way)
\begin{equation} \label{eq:ASEP_K}
  K(z) = \begin{pmatrix}
          \frac{(c-a)z^2+z}{cz^2+z-a} & \frac{c(z^2-1)}{cz^2+z-a} \\
          \frac{a(z^2-1)}{cz^2+z-a} & \frac{c-a+z}{cz^2+z-a}
         \end{pmatrix} \quad \mbox{and} \quad 
  \overline{K}(z) = \begin{pmatrix}
          \frac{(b-d)z^2-z}{bz^2-z-d} & \frac{b(z^2-1)}{bz^2-z-d} \\
          \frac{d(z^2-1)}{bz^2-z-d} & \frac{b-d-z}{bz^2-z-d}
         \end{pmatrix}.
\end{equation}
They satisfy the reflection equation \eqref{eq:reflection_equation} and the reversed reflection equation \eqref{eq:reflection_equation_reversed}
respectively. They also fulfill the Markovian \eqref{eq:markovian_K}, regularity \eqref{eq:regularity_K} and unitarity \eqref{eq:unitarity_K} properties.
They will be again used to define local Markovian operators acting on the first and last sites, which describe the interaction with particle reservoirs at the boundaries.
 
Following one more time the general procedure introduced in subsection \ref{subsec:parallel_update} we define the local Markovian operators using the 
$R$ and $K$ matrices
\begin{equation}
 U = \check R(\kappa^2), \qquad B=K(\kappa), \qquad \overline{B}=\overline{K}\left(\frac{1}{\kappa}\right). 
\end{equation}
It allows us to define an integrable discrete time Markov matrix defined by \eqref{eq:Markov_matrix_open}. We now give an explicit description of the 
local stochastic rules of the model.
 In the bulk the dynamics encoded by the matrix $U$ explicitly reads
\begin{equation}
 \begin{aligned}
  & 01 \longrightarrow 10 \quad \frac{(1-\kappa^2)t^2}{1-t^2\kappa^2}; \qquad  01 \longrightarrow 01 \quad \frac{1-t^2}{1-t^2\kappa^2}; \\
  & 10 \longrightarrow 01 \quad \frac{1-\kappa^2}{1-t^2\kappa^2}; \qquad  10 \longrightarrow 10 \quad \frac{\kappa^2(1-t^2)}{1-t^2\kappa^2}; \\
 \end{aligned}
\end{equation}
Note that the dynamics is asymmetric as announced: the particles jump to the left and to the right with different probabilities. It can for instance
describe a driving force in the bulk. Note that the SSEP dynamics is recovered by setting $\kappa\rightarrow e^{h\kappa}$, $t^2\rightarrow e^h$ 
and taking the limit $h$ going to zero.
On the left boundary the dynamics encoded by the matrix $B$ reads
\begin{equation}
 \begin{aligned}
  & 0 \longrightarrow 1 \quad \frac{a(1-\kappa^2)}{a-\kappa-c\kappa^2}; \qquad 0 \longrightarrow 0 \quad \frac{(a-c)\kappa^2-\kappa}{a-\kappa-c\kappa^2}; \\
  & 1 \longrightarrow 0 \quad \frac{c(1-\kappa^2)}{a-\kappa-c\kappa^2}; \qquad 1 \longrightarrow 1 \quad \frac{a-c-\kappa}{a-\kappa-c\kappa^2}; \\
 \end{aligned}
\end{equation}
It may for instance describe the interaction with a particle reservoir at fixed density.
On the right boundary the dynamics encoded by the matrix $\overline{B}$ reads
\begin{equation}
 \begin{aligned}
  & 0 \longrightarrow 1 \quad \frac{d(1-\kappa^2)}{b-\kappa-d\kappa^2}; \qquad 0 \longrightarrow 0 \quad \frac{b-d-\kappa}{b-\kappa-d\kappa^2}; \\
  & 1 \longrightarrow 0 \quad \frac{b(1-\kappa^2)}{b-\kappa-d\kappa^2}; \qquad 1 \longrightarrow 1 \quad \frac{(b-d)\kappa^2-\kappa}{b-\kappa-d\kappa^2}; \\
 \end{aligned}
\end{equation}
 Now that the discrete time dynamics has been introduced, we can look for its stationary distribution in a matrix product form following 
the general method presented in subsection \ref{subsec:stationary_state}. Again, we only need to construct a vector $\bm{A}(z)$ satisfying the 
ZF \eqref{eq:ZF} and GZ \eqref{eq:GZ} relations.

To achieve this goal, we define the algebraic-valued vector $\bm{A}(z)$ by the following expression 
\begin{equation} \label{eq:ASEP_A}
 \bm{A}(z) = \begin{pmatrix}
              z+\bm{e} \\ \frac{1}{z}+\bm{d}
             \end{pmatrix}.
\end{equation}
The generators $\bm{e}$ and $\bm{d}$ satisfy the following commutation relation
\begin{equation} \label{eq:ASEP_commutation_relation}
\bm{de}-t^2\bm{ed}=1-t^2
\end{equation}
and the relations on the boundary vectors $\bbra{W}$ and $\kket{V}$
\begin{equation} \label{eq:ASEP_relations_boundaries}
\bbra{W}\Big(a\bm{e}-c\bm{d}+1\Big)=0, \qquad \mbox{and} \qquad \Big(b\bm{d}-d\bm{e}+1\Big)\kket{V}=0.
\end{equation}
Let us stress the difference between $\bm{d}$ and $\bm{e}$ in bold, that are operators, and $a,b,c,d$ that are boundary parameters.
Note that these algebraic relations are essentially identical (up to some change of parameters) to the ones appearing when constructing 
the stationary state of the continuous time ASEP. It is known that
there exists an explicit representation of $\bm{e}$, $\bm{d}$, $\bbra{W}$ and $\kket{V}$, satisfying \eqref{eq:ASEP_commutation_relation} and 
\eqref{eq:ASEP_relations_boundaries}, as infinite dimensional matrices and vectors \cite{Sandow94,MartinRev}.
A direct computation shows that if the relations \eqref{eq:ASEP_commutation_relation} and \eqref{eq:ASEP_relations_boundaries} are fulfilled, then the 
vector $\bm{A}(z)$ satisfies the ZF \eqref{eq:ZF} and GZ \eqref{eq:GZ} relations (together with $\bbra{W}$ and $\kket{V}$).

Similarly to the symmetric case, we stress that, despite the similitude of the underlying algebraic relations, 
the stationary distribution of the present discrete time ASEP model is not the same as the one of the continuous time model because of 
the staggered construction\footnote{The stationary state of the continuous time ASEP is given by (with appropriate change of parameters)
$\steady = \frac{1}{Z_L}\bbra{W} \bm{A}(1) \otimes \cdots \otimes \bm{A}(1) \kket{V}$}
\begin{equation}
 \steady = \frac{1}{Z_L}\bbra{W} \bm{A}(\kappa) \otimes \bm{A}\left(\frac{1}{\kappa}\right) \otimes \bm{A}(\kappa) \otimes \cdots \otimes \bm{A}(\kappa) \kket{V}.
\end{equation}
We can in principle use this matrix product solution to compute physical quantities. However in the present case of the ASEP, we will not 
handle the computations up to the final result but rather give an idea on how to use the matrix product formalism  in this discrete time 
dynamics framework. Performing a complete derivation of the physical quantities would require dealing with the explicit representation of 
the algebraic elements $\bm{e}$, $\bm{d}$ , $\bbra{W}$ and $\kket{V}$ and introducing $q$-calculus and $q$-hypergeometric 
functions \cite{Sandow94,MartinRev} which is beyond the scope of this paper. 

The first step is to compute the normalization $Z_L$. We recall that it is defined as
\begin{equation}
 Z_L= Z_L(\kappa,1/\kappa,\dots,\kappa) = \bbra{W} \bm{C}(\kappa)\bm{C}(1/\kappa)\cdots \bm{C}(\kappa)\kket{V}
 = \bbra{W}\left(\kappa+\frac{1}{\kappa}+\bm{e}+\bm{d}\right)^L \kket{V}
\end{equation}
because $\bm{C}(z)=\bra{\sigma}\bm{A}(z)=\bm{C}(1/z)$. 
It depends explicitly on the parameter $\kappa$ and thus differs from the corresponding expression for the continuous time ASEP.
It would be interesting to compute the asymptotic behavior of this normalization for large system size $L$ and see how it is modified in 
comparison to the continuous time \cite{DEHP,Sandow94}. As we will see below the asymptotic behavior of the normalization plays an important role
in the determination of the phase diagram of the model.

We can also study the mean particle density on site $i$. Similarly to the symmetric case, taking into account the two steps Floquet dynamics, 
\begin{equation}
 \langle \tau_i \rangle = \frac{1}{Z_L}
 \bbra{W}\left(\kappa+\frac{1}{\kappa}+\bm{e}+\bm{d}\right)^{i-1}\left(\frac{1}{2}\left(\kappa+\frac{1}{\kappa}\right)+\bm{d}\right)\left(\kappa+\frac{1}{\kappa}+\bm{e}+\bm{d}\right)^{L-i}\kket{V}.
\end{equation}
We are finally interested in the mean particle current. We have first to compute the quantity
\begin{equation}
 \frac{1-\kappa^2}{1-t^2\kappa^2}\left(\frac{1}{\kappa}+\bm{d}\right)\left(\frac{1}{\kappa}+\bm{e}\right)
 -\frac{(1-\kappa^2)t^2}{1-t^2\kappa^2}(\kappa+\bm{e})(\kappa+\bm{d})
 =\left(\frac{1}{\kappa}-\kappa\right)\left(\kappa+\frac{1}{\kappa}+\bm{e}+\bm{d}\right),
\end{equation}
which is related to the mean number of particles jumping over a particular bond per time step.
Hence we deduce that
\begin{equation}
 \langle J \rangle = \left(\frac{1}{\kappa}-\kappa\right)\frac{Z_{L-1}}{Z_L}.
\end{equation}
It has a similar structure as the mean particle current of the continuous time ASEP but with an additional factor $1/\kappa-\kappa$.

\section{Integrable generalized exclusion processes from the fusion procedure. \label{sec:generalized_exclusion}}

This section is devoted to the construction of integrable generalized exclusion processes, \textit{i.e} models for which
the exclusion constraint of the SSEP and ASEP (there is at most one particle per site, $s=1$) is a bit relaxed: there is at most $s$ particles 
per site with $s>1$. The general idea to construct such models is the following. The $R$-matrix associated to the ASEP (respectively to the SSEP)
can be viewed as the fundamental representation (also known as spin $1/2$ representation) of the universal $R$-matrix associated to the 
quantum group $\cU_q(\hat{sl}_2)$ (respectively to the Yangian $\cY(\hat{sl}_2)$). We can construct the $R$-matrices associated to generalized 
exclusion processes by taking higher dimensional irreducible representations (\textit{i.e} higher spin representations) of the universal 
$R$-matrix. A well-known technique, called \textit{fusion procedure} \cite{Karowski,KulishS,KulishRS,Jimbo}, 
allows us to construct easily the higher dimensional representations of the 
$R$-matrix starting from the $R$-matrix in fundamental (spin $1/2$) representation. We summarize briefly the key points of the procedure in 
the next subsection \ref{subsec:fusion} because it will shed some light on the \textit{fused} matrix ansatz construction of subsection \ref{subsec:fused_MA}. 

\subsection{Fusion of R and K matrices and description of the new models. \label{subsec:fusion}}

We wish to give in this small paragraph some insight on the fusion procedure. It is not meant to be 
fully rigorous. The reader is invited to refer to \cite{Karowski,KulishS,KulishRS,Jimbo} and references therein or to the specific examples 
developed hereafter to get into the technical details.

Let us first introduce some notations and properties associated with quantum groups. The discussion holds for the quantum group $\cA=\cU_q(\hat{sl}_2)$,
corresponding to the asymmetric case, and also for $\cA = \cY(\hat{sl}_2)$ corresponding to the symmetric case.
The quantum group is endowed with an algebra homomorphism $\Delta:\cA \rightarrow \cA \otimes \cA$ called the \textit{coproduct}, which allows us to
construct easily tensor products of representations (see below).
There exists an element $\mathfrak{R} \in \cA \otimes \cA$ called \textit{universal R-matrix}, satisfying the Yang-Baxter equation
\begin{equation}
 \mathfrak{R}_{12}\mathfrak{R}_{13}\mathfrak{R}_{23}=\mathfrak{R}_{23}\mathfrak{R}_{13}\mathfrak{R}_{12}.
\end{equation}
It satisfies also convenient relations with the coproduct, such as $(1\otimes \Delta)(\mathfrak{R})=\mathfrak{R}_{12}\mathfrak{R}_{23}$.
The irreducible finite dimensional representations $\pi_z^{(m)}:\cA \rightarrow End(\cV_m)$ are labeled by a half-integer $m$ called spin, and depend 
on a spectral parameter $z$. The dimension of the representation is equal to $2m+1$. 
Taking finite dimensional representation of the element $\mathfrak{R}$
\begin{equation} \label{eq:R_spin_quelconque}
 (\pi_{z_1}^{(m)} \otimes \pi_{z_2}^{(l)})(\mathfrak{R}) := R^{(m,l)}\left(\frac{z_1}{z_2}\right)
\end{equation}
allows us to construct a finite dimensional matrix acting on $\cV_m \otimes \cV_l$ 
(\textit{i.e} on two sites with spin $m$ and $j$ respectively) satisfying the Yang-Baxter equation
\begin{equation}
 R^{(m,l)}_{12}\left(\frac{z_1}{z_2}\right)R^{(m,k)}_{13}\left(\frac{z_1}{z_3}\right)R^{(l,k)}_{23}\left(\frac{z_2}{z_3}\right)=
 R^{(l,k)}_{23}\left(\frac{z_2}{z_3}\right)R^{(m,k)}_{13}\left(\frac{z_1}{z_3}\right)R^{(m,l)}_{23}\left(\frac{z_1}{z_2}\right)
\end{equation}
Note that the R-matrix \eqref{eq:R_spin_quelconque} depends only on the ratio\footnote{This holds for $\cA=\cU_q(\hat{sl}_2)$. In the case 
$\cA = \cY(\hat{sl}_2)$ it instead depends on the difference.} $z_1/z_2$. The R-matrix \eqref{eq:ASEP_R} corresponds to the case $m=l=1/2$ and 
is equal to $R^{(1/2,1/2)}(z)$. 

The fusion procedure consists in constructing the matrix $R^{(m,l)}(z)$, for general half-integers $m$ and $l$, from $R^{(1/2,1/2)}(z)$, 
without having to deal with the universal R-matrix $\mathfrak{R}$ which is a rather complicated object in practice. 

For instance we would like to construct the representation $\pi_{z}^{(1)}$ from the tensor product of the fundamental representations
$\pi_{z_1}^{(1/2)}$ and $\pi_{z_2}^{(1/2)}$ which is defined with the help of the coproduct by 
$(\pi_{z_1}^{(1/2)} \otimes \pi_{z_2}^{(1/2)}) \Delta:\cA \rightarrow End(\cV_{1/2} \otimes \cV_{1/2})$. The latter tensor product representation 
is irreducible for generic spectral parameters $z_1$ and $z_2$ but it becomes reducible when the ratio $z_2/z_1=\mu$, with $\mu$ such that 
$R^{(1/2,1/2)}(\mu)$ is a projector. The representation $\pi_{z}^{(1)}$ is then obtained by projecting onto the invariant subspace, which is 
achieved by acting with the projector $R^{(1/2,1/2)}(\mu)$. It yields for instance the formula
\begin{eqnarray}
 R^{(1/2,1)}(z) & = & R^{(1/2,1/2)}(\mu).\Big[(\pi_{z}^{(1/2)} \otimes \pi_{\mu^{1/2}}^{(1/2)} \otimes \pi_{\mu^{-1/2}}^{(1/2)})(1 \otimes \Delta)(\mathfrak{R})\Big] .R^{(1/2,1/2)}(\mu)\\
 & = & R^{(1/2,1/2)}(\mu) R^{(1/2,1/2)}(z/\mu^{1/2}) R^{(1/2,1/2)}(z\mu^{1/2}) R^{(1/2,1/2)}(\mu).
\end{eqnarray}
Such kind of construction can be applied recursively to obtain $R^{(m,l)}(z)$, see for instance formulas 
\eqref{eq:GSSEP_half_fused_R} and \eqref{eq:GSSEP_R_definition} below.


A similar fusion procedure also exists to deal with open boundary conditions and construct higher dimensional $K$-matrices \cite{MezincescuN,FrappatNR}.

We illustrate more precisely these procedures by applying it to the SSEP and ASEP.

\paragraph{The symmetric case.}

We are first interested in the SSEP case. To start with the fusion procedure for this model, we need to identify a particular point $\mu$ at
which the $R$-matrix defined in \eqref{eq:SSEP_R} is a projector, \textit{i.e} $R(\mu)^2=R(\mu)$. It is straightforward to check that 
$\mu=1$ is the only point that fulfills this condition. It is then easy to realize that $R(1)$ is closely related to the rectangular matrices
\begin{equation}
 Q^{(l)} = \begin{pmatrix}
            1 & 0 & 0 & 0 \\
            0 & 1 & 1 & 0 \\
            0 & 0 & 0 & 1
           \end{pmatrix} \quad \mbox{and} \quad 
 Q^{(r)} = \begin{pmatrix}
            1 & 0 & 0 \\
            0 & 1/2 & 0 \\
            0 & 1/2 & 0 \\
            0 & 0 & 1
           \end{pmatrix}          
\end{equation}
through the following relations
\begin{equation} \label{eq:projectors_properties}
 Q^{(l)}Q^{(r)} = 1, \qquad Q^{(r)}Q^{(l)} = R(1), \qquad Q^{(l)}R(1)=Q^{(l)}, \qquad R(1)Q^{(r)}=Q^{(r)}.
\end{equation}
The rectangular matrices $Q^{(l)}$ and $Q^{(r)}$ will allow us to project on the invariant subspace of the tensor representation (going from a 
dimension $4$ space, obtained as the tensor product of two fundamental representations of dimension $2$, to a dimension $3$ irreducible subspace).
More precisely we can fuse the second space of the $R$-matrix as follows
\begin{equation} \label{eq:GSSEP_half_fused_R}
 R_{i,<jk>}(z)= Q^{(l)}_{jk} R_{ij}\left(z-\frac{1}{2}\right) R_{ik}\left(z+\frac{1}{2}\right)Q^{(r)}_{jk}.
\end{equation}
The resulting matrix $R_{i,<jk>}(z)$ acts on $\CC^2 \otimes \CC^3$ and satisfies some specific Yang-Baxter equation, see \cite{FrappatNR}.
We are now left with the fusion of the first space
\begin{equation} \label{eq:GSSEP_R_definition}
 \cR(z)=R_{<hi>,<jk>}(z)= Q^{(l)}_{hi} R_{h,<jk>}\left(z+\frac{1}{2}\right) R_{i,<jk>}\left(z-\frac{1}{2}\right)Q^{(r)}_{hi}.
\end{equation} 
The matrix $\cR(z)$ acts on $\CC^3 \otimes \CC^3$ and will correspond to a process with at most $2$ particles allowed on each site (\textup{i.e} it
corresponds to $s=2$). It has the following explicit expression, given in the basis $\ket{0}\otimes\ket{0}$, $\ket{0}\otimes\ket{1}$, 
$\ket{0}\otimes\ket{2}$, $\ket{1}\otimes\ket{0}$, $\ket{1}\otimes\ket{1}$, $\ket{1}\otimes\ket{2}$, $\ket{2}\otimes\ket{0}$, $\ket{2}\otimes\ket{1}$, 
$\ket{2}\otimes\ket{2}$ (ordered this way)
\begin{equation} \label{eq:GSSEP_R}
 \cR(z) =  \begin{pmatrix}
            1 & 0 & 0 & 0 & 0 & 0 & 0 & 0 & 0 \\
            0 & \frac{z}{z+2} & 0 & \frac{2}{z+2} & 0 & 0 & 0 & 0 & 0 \\
            0 & 0 & \frac{z(z-1)}{(z+1)(z+2)} & 0 & \frac{z}{(z+1)(z+2)} & 0 & \frac{2}{(z+1)(z+2)} & 0 & 0 \\
            0 & \frac{2}{z+2} & 0 & \frac{z}{z+2} & 0 & 0 & 0 & 0 & 0 \\
            0 & 0 & \frac{4z}{(z+1)(z+2)} & 0 & \frac{z^2+z+2}{(z+1)(z+2)} & 0 & \frac{4z}{(z+1)(z+2)} & 0 & 0 \\
            0 & 0 & 0 & 0 & 0 & \frac{z}{z+2} & 0 & \frac{2}{z+2} & 0 \\
            0 & 0 & \frac{2}{(z+1)(z+2)} & 0 & \frac{z}{(z+1)(z+2)} & 0 & \frac{z(z-1)}{(z+1)(z+2)} & 0 & 0 \\
            0 & 0 & 0 & 0 & 0 & \frac{2}{z+2} & 0 & \frac{z}{z+2} & 0 \\
            0 & 0 & 0 & 0 & 0 & 0 & 0 & 0 & 1 
           \end{pmatrix}
\end{equation}
Using the fact that the matrix $R(z)$ satisfies the Yang-Baxter equation \eqref{eq:YBE_additive} and also 
the properties \eqref{eq:projectors_properties}, it is possible to show that $\cR(z)$ satisfies the Yang-Baxter equation\footnote{It can also
be checked directly using the explicit expression \eqref{eq:GSSEP_R}.} 
\begin{equation}
  \cR_{12}\left(z_1-z_2\right) \cR_{13}\left(z_1-z_3\right) \cR_{23}\left(z_2-z_3\right) = 
 \cR_{23}\left(z_2-z_3\right) \cR_{13}\left(z_1-z_3\right) \cR_{12}\left(z_1-z_2\right)  
  \end{equation}
It is also straightforward to show, using either the definition \eqref{eq:GSSEP_R_definition} or the explicit expression \eqref{eq:GSSEP_R}, that 
the matrix $\cR(z)$ satisfies the Markovian, regularity and unitarity properties. 
 A similar fusion procedure can also be applied to the $K$-matrices \cite{FrappatNR}, defining
\begin{equation}
 \cK(z) = K_{<ij>}(z) = Q^{(l)}_{ij} K_i\left(z-\frac{1}{2}\right)R_{ji}(2z)K_j\left(z+\frac{1}{2}\right)Q^{(r)}_{ij}
\end{equation}
and 
\begin{equation}
 \overline{\cK}(z) = \overline{K}_{<ij>}(z) = Q^{(l)}_{ij} \overline{K}_i\left(z-\frac{1}{2}\right)R_{ji}(2z)^{-1}\overline{K}_j\left(z+\frac{1}{2}\right)Q^{(l)}_{ij}.
\end{equation}
The matrices $\cK(z)$ and $\overline{\cK}(z)$ both act on $\CC^3$ and take the following explicit expressions, written in 
the basis $\ket{0}$, $\ket{1}$ and $\ket{2}$
\begin{eqnarray} \label{eq:GSSEP_K}
 & & \cK(z) = \\
 & &  \hspace{-5mm}    \begin{pmatrix}
           * & \frac{4cz\big[(2z-1)(c-a)+2\big]}{\big[(2z-1)(a+c)+2\big]\big[(2z+1)(a+c)+2\big]} & \frac{8c^2z(2z-1)}{\big[(2z-1)(a+c)+2\big]\big[(2z+1)(a+c)+2\big]} \\
           \frac{8az\big[(2z-1)(c-a)+2\big]}{\big[(2z-1)(a+c)+2\big]\big[(2z+1)(a+c)+2\big]} & * & \frac{8cz\big[(2z-1)(a-c)+2\big]}{\big[(2z-1)(a+c)+2\big]\big[(2z+1)(a+c)+2\big]} \\
           \frac{8a^2z(2z-1)}{\big[(2z-1)(a+c)+2\big]\big[(2z+1)(a+c)+2\big]} & \frac{4az\big[(2z-1)(a-c)+2\big]}{\big[(2z-1)(a+c)+2\big]\big[(2z+1)(a+c)+2\big]} & *
          \end{pmatrix} \nonumber 
\end{eqnarray}
and 
\begin{eqnarray} \label{eq:GSSEP_Kb}
 & & \overline{\cK}(z) = \\
 & &  \hspace{-5mm}    \begin{pmatrix}
           * & \frac{4bz\big[(2z+1)(b-d)-2\big]}{\big[(2z-1)(b+d)-2\big]\big[(2z+1)(b+d)-2\big]} & \frac{8b^2z(2z+1)}{\big[(2z-1)(b+d)-2\big]\big[(2z+1)(b+d)-2\big]} \\
           \frac{8dz\big[(2z+1)(b-d)-2\big]}{\big[(2z-1)(b+d)-2\big]\big[(2z+1)(b+d)-2\big]} & * & \frac{8bz\big[(2z+1)(d-b)-2\big]}{\big[(2z-1)(b+d)-2\big]\big[(2z+1)(b+d)-2\big]} \\
           \frac{8d^2z(2z+1)}{\big[(2z-1)(b+d)-2\big]\big[(2z+1)(b+d)-2\big]} & \frac{4dz\big[(2z+1)(d-b)-2\big]}{\big[(2z-1)(b+d)-2\big]\big[(2z+1)(b+d)-2\big]} & * 
          \end{pmatrix}. \nonumber 
\end{eqnarray}
Note that in order to make these matrices easier to read we have replaced the diagonal entries (which do not enjoy a factorized form) by $*$. 
We also have used $[\cdot ]$ to denote usual brackets. 
The value of the diagonal entries can be easily deduced from the rest of the matrix using the Markovian property: the sum of the entries on each 
column is equal to one. The regularity and unitarity properties are also fulfilled.
These matrices satisfy the reflection equations (it can be checked from the explicit expressions \eqref{eq:GSSEP_K} and \eqref{eq:GSSEP_Kb})
\begin{equation} 
\cR_{12}\left(z_1-z_2\right) \cK_1(z_1) \cR_{21}(z_1+z_2) \cK_2(z_2) =
  \cK_2(z_2) \cR_{12}(z_1+z_2) \cK_1(z_1) \cR_{21} \left(z_1-z_2\right)  
\end{equation}
and
\begin{equation} 
  \cR_{12}(z_1-z_2)^{-1} \overline{\cK}_1(z_1) R_{21}(z_1+z_2)^{-1} \overline{\cK}_2(z_2) =
  \overline{\cK}_2(z_2) \cR_{12}(z_1+z_2)^{-1} \overline{\cK}_1(z_1) \cR_{21}(z_1-z_2)^{-1}.
\end{equation}

We are now equipped to describe the dynamics of the model, \textit{i.e} the transition probabilities between the different configurations. 
We recall that the local stochastic dynamics is encoded in the following matrices
\begin{equation}
 \cU = \check R (2\kappa), \qquad \cB = \cK(\kappa), \qquad \overline{\cB} = \overline{\cK}(-\kappa).
\end{equation}
As explained in subsection \ref{subsec:parallel_update} the stochastic dynamics on the whole lattice is then defined by the Markov matrix
\begin{equation}
 M = \UU^{e}\UU^{o} = \left(\cB_1 \prod_{k=1}^{\frac{L-1}{2}} \cU_{2k,2k+1}\right) \left(\overline{\cB}_L \prod_{k=1}^{\frac{L-1}{2}} \cU_{2k-1,2k}\right).
\end{equation}

We now give a precise description of the local stochastic rules.
In the bulk the dynamics encoded by the matrix $\cU$ explicitly reads
\begin{equation}
 \begin{aligned}
 & 01 \longrightarrow 10 \quad \frac{\kappa}{\kappa+1}; \qquad 01 \longrightarrow 01 \quad \frac{1}{\kappa+1}; \\
 & 10 \longrightarrow 01 \quad \frac{\kappa}{\kappa+1}; \qquad 10 \longrightarrow 10 \quad \frac{1}{\kappa+1}; \\
 & 12 \longrightarrow 21 \quad \frac{\kappa}{\kappa+1}; \qquad 12 \longrightarrow 12 \quad \frac{1}{\kappa+1}; \\
 & 21 \longrightarrow 12 \quad \frac{\kappa}{\kappa+1}; \qquad 21 \longrightarrow 21 \quad \frac{1}{\kappa+1}; \\
 & 02 \longrightarrow 11 \quad \frac{4\kappa}{(2\kappa+1)(\kappa+1)}; \qquad 02 \longrightarrow 20 \quad \frac{(2\kappa-1)\kappa}{(2\kappa+1)(\kappa+1)};
  \qquad 02 \longrightarrow 02 \quad \frac{1}{(2\kappa+1)(\kappa+1)}; \\
 & 20 \longrightarrow 11 \quad \frac{4\kappa}{(2\kappa+1)(\kappa+1)}; \qquad 20 \longrightarrow 02 \quad \frac{(2\kappa-1)\kappa}{(2\kappa+1)(\kappa+1)};
  \qquad 20 \longrightarrow 20 \quad \frac{1}{(2\kappa+1)(\kappa+1)}; \\
 & 11 \longrightarrow 02 \quad \frac{\kappa}{(2\kappa+1)(\kappa+1)}; \qquad 11 \longrightarrow 20 \quad \frac{\kappa}{(2\kappa+1)(\kappa+1)};
 \qquad 11 \longrightarrow 11 \quad \frac{2\kappa^2+\kappa+1}{(2\kappa+1)(\kappa+1)};
 \end{aligned}
\end{equation}
Note that we can observe a left-right symmetry in the transition probabilities. It is explained by the fact that $\cU=\cP \cU \cP$, where 
$\cP$ is the permutation operator in $\CC^3 \otimes \CC^3$. We can also observe that the transition probabilities are not simply proportional 
to the number of particles lying on the departure site but depend in a non-trivial way on the composition of both departure and arrival sites. 
It thus describes non trivial interactions between the particles.
On the left boundary the dynamics encoded by the matrix $\cB$ reads
\begin{equation}
 \begin{aligned}
  & \hspace{-1cm} 0 \longrightarrow 1 \quad \frac{8a\kappa\big[(2\kappa-1)(c-a)+2\big]}{\big[(2\kappa-1)(a+c)+2\big]\big[(2\kappa+1)(a+c)+2\big]}; \qquad 
  0 \longrightarrow 2 \quad \frac{8a^2\kappa(2\kappa-1)}{\big[(2\kappa-1)(a+c)+2\big]\big[(2\kappa+1)(a+c)+2\big]}; \\
  & \hspace{-1cm} 1 \longrightarrow 0 \quad \frac{4c\kappa\big[(2\kappa-1)(c-a)+2\big]}{\big[(2\kappa-1)(a+c)+2\big]\big[(2\kappa+1)(a+c)+2\big]}; \qquad
  1 \longrightarrow 2 \quad \frac{4a\kappa\big[(2\kappa-1)(a-c)+2\big]}{\big[(2\kappa-1)(a+c)+2\big]\big[(2\kappa+1)(a+c)+2\big]}; \\
  & \hspace{-1cm} 2 \longrightarrow 0 \quad \frac{8c^2\kappa(2\kappa-1)}{\big[(2\kappa-1)(a+c)+2\big]\big[(2\kappa+1)(a+c)+2\big]}; \qquad 
  2 \longrightarrow 1 \quad \frac{8c\kappa\big[(2\kappa-1)(a-c)+2\big]}{\big[(2\kappa-1)(a+c)+2\big]\big[(2\kappa+1)(a+c)+2\big]}; 
 \end{aligned}
\end{equation}
It describes a non-trivial interaction with a particle reservoir.
On the right boundary the dynamics encoded by the matrix $\overline{\cB}$ reads
\begin{equation}
 \begin{aligned}
  & \hspace{-1cm} 0 \longrightarrow 1 \quad \frac{8d\kappa\big[(2\kappa-1)(b-d)+2\big]}{\big[(2\kappa-1)(b+d)+2\big]\big[(2\kappa+1)(b+d)+2\big]}; \qquad 
  0 \longrightarrow 2 \quad \frac{8d^2\kappa(2\kappa-1)}{\big[(2\kappa-1)(b+d)+2\big]\big[(2\kappa+1)(b+d)+2\big]}; \\
  & \hspace{-1cm} 1 \longrightarrow 0 \quad \frac{4b\kappa\big[(2\kappa-1)(b-d)+2\big]}{\big[(2\kappa-1)(b+d)+2\big]\big[(2\kappa+1)(b+d)+2\big]}; \qquad
  1 \longrightarrow 2 \quad \frac{4d\kappa\big[(2\kappa-1)(d-b)+2\big]}{\big[(2\kappa-1)(b+d)+2\big]\big[(2\kappa+1)(b+d)+2\big]}; \\
  & \hspace{-1cm} 2 \longrightarrow 0 \quad \frac{8b^2\kappa(2\kappa-1)}{\big[(2\kappa-1)(b+d)+2\big]\big[(2\kappa+1)(b+d)+2\big]}; \qquad 
  2 \longrightarrow 1 \quad \frac{8b\kappa\big[(2\kappa-1)(d-b)+2\big]}{\big[(2\kappa-1)(b+d)+2\big]\big[(2\kappa+1)(b+d)+2\big]}; 
 \end{aligned}
\end{equation}

\begin{remark}
 The dynamics of the model can be considerably simplified by considering the particular value $\kappa=1/2$. The transitions involving the
 simultaneous jumps of two particles have in this case a vanishing probability.   
\end{remark}

As far as we know, this model is the first example of integrable generalized exclusion process with open boundary conditions. It emphasizes the 
width of applications of the generic procedure presented in subsection \ref{subsec:parallel_update} to construct discrete time models.

\paragraph{The asymmetric case.}

We are now interested in the ASEP case. To apply again the fusion procedure on this model, we need to identify the particular point $\mu$ at
which the $R$-matrix defined in \eqref{eq:ASEP_R} is a projector. It is possible to establish that 
$\mu=t^2$ is the only point that fulfills this condition. It is then easy to realize that $R(t^2)$ is closely related to the rectangular matrices

\begin{equation}
 Q^{(l)} = \begin{pmatrix}
            1 & 0 & 0 & 0 \\
            0 & 1 & 1 & 0 \\
            0 & 0 & 0 & 1
           \end{pmatrix} \quad \mbox{and} \quad 
 Q^{(r)} = \begin{pmatrix}
            1 & 0 & 0 \\
            0 & t^2/(1+t^2) & 0 \\
            0 & 1/(1+t^2) & 0 \\
            0 & 0 & 1
           \end{pmatrix}          
\end{equation}
through similar relations as in the symmetric case
\begin{equation} \label{eq:projectors_properties_bis}
 Q^{(l)}Q^{(r)} = 1_3, \qquad Q^{(r)}Q^{(l)} = R(t^2), \qquad Q^{(l)}R(t^2)=Q^{(l)}, \qquad R(t^2)Q^{(r)}=Q^{(r)}.
\end{equation}
The rectangular matrices $Q^{(l)}$ and $Q^{(r)}$ will again allow us to project on the invariant subspace of the tensor representation.
Similarly to the symmetric case, we can fuse the second space of the $R$-matrix as follows
\begin{equation}
 R_{i,<jk>}(z)= Q^{(l)}_{jk} R_{ij}\left(\frac{z}{t}\right) R_{ik}(zt)Q^{(r)}_{jk}.
\end{equation}
The matrix $R_{i,<jk>}(z)$ acts on $\CC^2 \otimes \CC^3$ and satisfies also a Yang-Baxter equation. 
We are then left with the fusion of the first space
\begin{equation}
 \cR(z)=R_{<hi>,<jk>}(z)= Q^{(l)}_{hi} R_{h,<jk>}\left(zt\right) R_{i,<jk>}\left(\frac{z}{t}\right)Q^{(r)}_{hi}.
\end{equation}
The matrix $\cR(z)$ acts on $\CC^3 \otimes \CC^3$ and will correspond again to a process with at most $2$ particles allowed on each site (\textup{i.e} it
corresponds to $s=2$). It has the following explicit expression, given in the basis $\ket{0}\otimes\ket{0}$, $\ket{0}\otimes\ket{1}$, 
$\ket{0}\otimes\ket{2}$, $\ket{1}\otimes\ket{0}$, $\ket{1}\otimes\ket{1}$, $\ket{1}\otimes\ket{2}$, $\ket{2}\otimes\ket{0}$, $\ket{2}\otimes\ket{1}$, 
$\ket{2}\otimes\ket{2}$ (ordered this way)
\begin{eqnarray} \label{eq:GASEP_R}
 & & \cR(z) = \\
 & & \hspace{-12mm} \begin{pmatrix}
            1 & 0 & 0 & 0 & 0 & 0 & 0 & 0 & 0 \\
            0 & \frac{t^4(1-z)}{1-zt^4} & 0 & \frac{z(1-t^4)}{1-zt^4} & 0 & 0 & 0 & 0 & 0 \\
            0 & 0 & \frac{t^6(t^2-z)(1-z)}{(1-zt^2)(1-zt^4)} & 0 & \frac{zt^4(1-t^2)(1-z)}{(1-zt^2)(1-zt^4)} & 0 & \frac{z^2(1-t^2)(1-t^4)}{(1-zt^2)(1-zt^4)} & 0 & 0 \\
            0 & \frac{1-t^4}{1-zt^4} & 0 & \frac{1-z}{1-zt^4} & 0 & 0 & 0 & 0 & 0 \\
            0 & 0 & \frac{t^2(1+t^2)(1-t^4)(1-z)}{(1-zt^2)(1-zt^4)} & 0 & \frac{zt^6+z^2t^4-2zt^4-2zt^2+t^2+z}{(1-zt^2)(1-zt^4)} & 0 & \frac{z(1-z)(1+t^2)(1-t^4)}{t^2(1-zt^2)(1-zt^4)} & 0 & 0 \\
            0 & 0 & 0 & 0 & 0 & \frac{t^4(1-z)}{1-zt^4} & 0 & \frac{z(1-t^4)}{1-zt^4} & 0 \\
            0 & 0 & \frac{(1-t^2)(1-t^4)}{(1-zt^2)(1-zt^4)} & 0 & \frac{(1-z)(1-t^2)}{(1-zt^2)(1-zt^4)} & 0 & \frac{(t^2-z)(1-z)}{t^2(1-zt^2)(1-zt^4)} & 0 & 0 \\
            0 & 0 & 0 & 0 & 0 & \frac{1-t^4}{1-zt^4} & 0 & \frac{1-z}{1-zt^4} & 0 \\
            0 & 0 & 0 & 0 & 0 & 0 & 0 & 0 & 1 
           \end{pmatrix} \nonumber 
\end{eqnarray}
Using the fact that the matrix $R(z)$ satisfies the Yang-Baxter equation \eqref{eq:YBE} and also 
the properties \eqref{eq:projectors_properties_bis}, it is possible to show that $\cR(z)$ satisfies also the Yang-Baxter equation 
\begin{equation}
  \cR_{12}\left(\frac{z_1}{z_2}\right) \cR_{13}\left(\frac{z_1}{z_3}\right) \cR_{23}\left(\frac{z_2}{z_3}\right) = 
 \cR_{23}\left(\frac{z_2}{z_3}\right) \cR_{13}\left(\frac{z_1}{z_3}\right) \cR_{12}\left(\frac{z_1}{z_2}\right).  
  \end{equation}
 A direct computation using for instance the explicit expression \eqref{eq:GASEP_R}, shows that 
the matrix $\cR(z)$ satisfies the Markovian, regularity and unitarity properties.
A similar fusion procedure can also be applied to the $K$-matrices \cite{FrappatNR}, defining
\begin{equation}
 \cK(z) = K_{<ij>}(z) = Q^{(l)}_{ij} K_i\left(\frac{z}{t}\right)R_{ji}(z^2)K_j\left(zt\right) Q^{(r)}_{ij}
\end{equation}
and 
\begin{equation}
 \overline{\cK}(z) = \overline{K}_{<ij>}(z) = Q^{(l)}_{ij} \overline{K}_i\left(\frac{z}{t}\right)R_{ji}(z^2)^{-1}\overline{K}_j\left(zt\right)Q^{(r)}_{ij}.
\end{equation}
The matrices $\cK(z)$ and $\overline{\cK}(z)$ both act on $\CC^3$ and they take the following explicit expressions, written in 
the basis $\ket{0}$, $\ket{1}$ and $\ket{2}$
\begin{equation} \label{eq:GASEP_K}
\cK(z) =  \begin{pmatrix}
           * & \frac{ct^2z(1-z^2)(az-cz-t)}{(at^2-cz^2-zt)(a-ct^2z^2-zt)} & \frac{c^2t^2(t^2-z^2)(1-z^2)}{(at^2-cz^2-zt)(a-ct^2z^2-zt)} \\
           \frac{a(1+t^2)z(1-z^2)(az-cz-t)}{(at^2-cz^2-zt)(a-ct^2z^2-zt)} & * & \frac{ct(1+t^2)(1-z^2)(at-ct-z)}{(at^2-cz^2-zt)(a-ct^2z^2-zt)} \\
           \frac{a^2(t^2-z^2)(1-z^2)}{(at^2-cz^2-zt)(a-ct^2z^2-zt)} & \frac{at(1-z^2)(at-ct-z)}{(at^2-cz^2-zt)(a-ct^2z^2-zt)} & * 
          \end{pmatrix} 
\end{equation}
and 
\begin{equation} \label{eq:GASEP_Kb}
\overline{\cK}(z) = \begin{pmatrix}
           * & \frac{btz(1-z^2)(1+dzt-bzt)}{(bz^2-dt^2-zt)(bt^2z^2-d-zt)} & \frac{b^2(1-z^2)(1-t^2z^2)}{(bz^2-dt^2-zt)(bt^2z^2-d-zt)} \\
           \frac{dt(1+t^2)z(1-z^2)(1+dzt-bzt)}{(bz^2-dt^2-zt)(bt^2z^2-d-zt)} & * & \frac{b(1+t^2)(1-z^2)(d-b+zt)}{(bz^2-dt^2-zt)(bt^2z^2-d-zt)} \\
           \frac{d^2t^2(1-z^2)(1-t^2z^2)}{(bz^2-dt^2-zt)(bt^2z^2-d-zt)} & \frac{dt^2(1-z^2)(d-b+zt)}{(bz^2-dt^2-zt)(bt^2z^2-d-zt)} & * 
          \end{pmatrix}. 
\end{equation}
Note that in order to make these matrices easier to read we have replaced the diagonal entries (which do not enjoy a factorized form) by $*$. 
The value of the diagonal entries can be easily deduced from the rest of the matrix using the Markovian property: the sum of the entries on each 
column is equal to one. The regularity and unitarity properties are also satisfied.
From the explicit expressions \eqref{eq:GASEP_K} and \eqref{eq:GASEP_Kb}, it can be checked that the $K$-matrices fulfill the reflection equations
\begin{equation} 
\cR_{12}\left(\frac{z_1}{z_2}\right) \cK_1(z_1) \cR_{21}(z_1 z_2) \cK_2(z_2) =
  \cK_2(z_2) \cR_{12}(z_1 z_2) \cK_1(z_1) \cR_{21} \left(\frac{z_1}{z_2}\right)  
\end{equation}
and
\begin{equation} 
  \cR_{12}\left(\frac{z_1}{z_2}\right)^{-1} \overline{\cK}_1(z_1) R_{21}(z_1 z_2)^{-1} \overline{\cK}_2(z_2) =
  \overline{\cK}_2(z_2) \cR_{12}(z_1 z_2)^{-1} \overline{\cK}_1(z_1) \cR_{21}\left(\frac{z_1}{z_2}\right)^{-1}.
\end{equation}

We are now equipped to describe the dynamics of the model, \textit{i.e} the transition probabilities between the different configurations. 
We recall that the local stochastic dynamics is encoded in the following matrices
\begin{equation}
 \cU = \check R (\kappa^2), \qquad \cB = \cK(\kappa), \qquad \overline{\cB} = \overline{\cK}\left(\frac{1}{\kappa}\right).
\end{equation}
We recall also that the stochastic dynamics on the whole lattice is then defined by the Markov matrix
\begin{equation}
 M = \UU^{e}\UU^{o} = \left(\cB_1 \prod_{k=1}^{\frac{L-1}{2}} \cU_{2k,2k+1}\right) \left(\overline{\cB}_L \prod_{k=1}^{\frac{L-1}{2}} \cU_{2k-1,2k}\right).
\end{equation}

We now give a precise description of the local stochastic rules.
In the bulk the dynamics encoded by the matrix $\cU$ explicitly reads
\begin{equation}
 \begin{aligned}
 & 01 \longrightarrow 10 \quad \frac{t^4(1-\kappa^2)}{1-t^4\kappa^2}; \qquad 01 \longrightarrow 01 \quad \frac{1-t^4}{1-t^4\kappa^2}; \\
 & 10 \longrightarrow 01 \quad \frac{1-\kappa^2}{1-t^4\kappa^2}; \qquad 10 \longrightarrow 10 \quad \frac{\kappa^2(1-t^2)}{1-t^4\kappa^2}; \\
 & 12 \longrightarrow 21 \quad \frac{t^4(1-\kappa^2)}{1-t^4\kappa^2}; \qquad 12 \longrightarrow 12 \quad \frac{1-t^4}{1-t^4\kappa^2}; \\
 & 21 \longrightarrow 12 \quad \frac{1-\kappa^2}{1-t^4\kappa^2}; \qquad 21 \longrightarrow 21 \quad \frac{\kappa^2(1-t^4)}{1-t^4\kappa^2}; \\
 & 02 \longrightarrow 11 \quad \frac{t^2(1+t^2)(1-t^4)(1-\kappa^2)}{(1-t^2\kappa^2)(1-t^4\kappa^2)}; \qquad 
 02 \longrightarrow 20 \quad \frac{t^6(t^2-\kappa^2)(1-\kappa^2)}{(1-t^2\kappa^2)(1-t^4\kappa^2)};\\
 & 20 \longrightarrow 11 \quad \frac{(1+t^2)(1-t^4)\kappa^2(1-\kappa^2)}{t^2(1-t^2\kappa^2)(1-t^4\kappa^2)}; \qquad 
 20 \longrightarrow 02 \quad \frac{(t^2-\kappa^2)(1-\kappa^2)}{t^2(1-t^2\kappa^2)(1-t^4\kappa^2)}; \\
 & 11 \longrightarrow 02 \quad \frac{(1-t^2)(1-\kappa^2)}{(1-t^2\kappa^2)(1-t^4\kappa^2)}; \qquad 
 11 \longrightarrow 20 \quad \frac{t^4\kappa^2(1-t^2)(1-\kappa^2)}{(1-t^2\kappa^2)(1-t^4\kappa^2)};
 \end{aligned}
\end{equation}
Note that the left-right symmetry in the bulk dynamics is broken. It can describe the driving of an external force. Moreover the 
transition probabilities depend in a complex way on the number of particles on the departure and arrival sites, and suggest a non-trivial interaction 
between the particles.
On the left boundary the dynamics encoded by the matrix $\cB$ reads
\begin{equation}
 \begin{aligned}
  & 0 \longrightarrow 1 \quad \frac{a(1+t^2)\kappa(1-\kappa^2)(a\kappa-c\kappa-t)}{(at^2-c\kappa^2-\kappa t)(a-ct^2\kappa^2-\kappa t)}; \qquad 
  0 \longrightarrow 2 \quad \frac{a^2(t^2-\kappa^2)(1-\kappa^2)}{(at^2-c\kappa^2-\kappa t)(a-ct^2\kappa^2-\kappa t)}; \\
  & 1 \longrightarrow 0 \quad \frac{ct^2\kappa(1-\kappa^2)(a\kappa-c\kappa-t)}{(at^2-c\kappa^2-\kappa t)(a-ct^2\kappa^2-\kappa t)}; \qquad
  1 \longrightarrow 2 \quad \frac{at(1-\kappa^2)(at-ct-\kappa)}{(at^2-c\kappa^2-\kappa t)(a-ct^2\kappa^2-\kappa t)}; \\
  & 2 \longrightarrow 0 \quad \frac{c^2t^2(t^2-\kappa^2)(1-\kappa^2)}{(at^2-c\kappa^2-\kappa t)(a-ct^2\kappa^2-\kappa t)}; \qquad 
  2 \longrightarrow 1 \quad \frac{ct(1+t^2)(1-\kappa^2)(at-ct-\kappa)}{(at^2-c\kappa^2-\kappa t)(a-ct^2\kappa^2-\kappa t)}; 
 \end{aligned}
\end{equation}
It describes a non-trivial interaction with a particle reservoir.
On the right boundary the dynamics encoded by the matrix $\overline{\cB}$ reads
\begin{equation}
 \begin{aligned}
  & 0 \longrightarrow 1 \quad \frac{dt(1+t^2)(1-\kappa^2)(bt-dt-\kappa)}{(bt^2-d\kappa^2-\kappa t)(b-dt^2\kappa^2-\kappa t)}; \qquad 
  0 \longrightarrow 2 \quad \frac{d^2t^2(t^2-\kappa^2)(1-\kappa^2)}{(bt^2-d\kappa^2-\kappa t)(b-dt^2\kappa^2-\kappa t)}; \\
  & 1 \longrightarrow 0 \quad \frac{bt(1-\kappa^2)(bt-dt-\kappa)}{(bt^2-d\kappa^2-\kappa t)(b-dt^2\kappa^2-\kappa t)}; \qquad
  1 \longrightarrow 2 \quad \frac{dt^2\kappa(1-\kappa^2)(b\kappa-d\kappa-t)}{(bt^2-d\kappa^2-\kappa t)(b-dt^2\kappa^2-\kappa t)}; \\
  & 2 \longrightarrow 0 \quad \frac{b^2(t^2-\kappa^2)(1-\kappa^2)}{(bt^2-d\kappa^2-\kappa t)(b-dt^2\kappa^2-\kappa t)}; \qquad 
  2 \longrightarrow 1 \quad \frac{b(1+t^2)\kappa(1-\kappa^2)(b\kappa-d\kappa-t)}{(bt^2-d\kappa^2-\kappa t)(b-dt^2\kappa^2-\kappa t)}; 
 \end{aligned}
\end{equation}

\begin{remark}
 The dynamics of the model can be considerably simplified by considering the particular value $\kappa=t$. The transitions involving the
 simultaneous jumps of two particles have in this case a vanishing probability.   
\end{remark}

\subsection{Fused matrix ansatz. \label{subsec:fused_MA}}

We introduce in this subsection a new method to construct explicitly the stationary state of the (symmetric and asymmetric) 
generalized exclusion processes in a matrix product form, using the matrices involved in the solution of the strict exclusion case 
(when there is at most one particle per site, \textit{i.e} for $s=1$). We call this method, which is heavily inspired by the fusion procedure,
``fused'' matrix ansatz. We apply it to the generalized exclusion processes (where at most two particles are allowed on the same site, 
\textit{i.e} where $s=2$) introduced in subsection \ref{subsec:fusion}. We stress nevertheless that the fused matrix ansatz technique appears 
as quite general and may be adapted to solve also higher dimensional models (\textit{i.e} for $s>2$). 

\paragraph{The symmetric case.}

Following the approach presented in subsection \ref{subsec:stationary_state}, the goal is to construct a vector $\bm{\cA}(z)$ with algebraic entries
such that the ZF and GZ relations are fulfilled. The stationary state is then easily constructed from this vector. For such a purpose we take 
advantage of the particular construction of the fused matrices $\cR$, $\cK$ and $\overline{\cK}$ and of the fact that we already know a vector
that fulfill the ZF and GZ equations associated to the matrices $R$, $K$ and $\overline{K}$ for the SSEP (\textit{i.e} for $s=1$).
The idea is to mimic the fusion procedure by defining
\begin{equation} \label{eq:GSSEP_fused_matrix_ansatz}
 \bm{\cA}(z) = Q^{(l)} \bm{A}\left(z-\frac{1}{2}\right) \otimes \bm{A}\left(z+\frac{1}{2}\right)
\end{equation}
where $\bm{A}(z)$ is the vector, defined in \eqref{eq:SSEP_A}, used to construct the stationary state of the SSEP (with at most one particle per site,
\textit{i.e} for $s=1$).
Writing it out explicitly it has the following expression
\begin{equation} \label{eq:GSSEP_A}
 \bm{\cA}(z) = \begin{pmatrix}
                \left(z+\frac{1}{2}\right)\left(z-\frac{1}{2}\right)-2z\bm{E}+\bm{E}^2 \\ 
                -2\left(z+\frac{1}{2}\right)\left(z-\frac{1}{2}\right)+2z(\bm{E}-\bm{D})+\bm{ED}+\bm{DE} \\
                \left(z+\frac{1}{2}\right)\left(z-\frac{1}{2}\right)+2z\bm{D}+\bm{D}^2
               \end{pmatrix}
\end{equation}
From expression \eqref{eq:GSSEP_fused_matrix_ansatz}, a direct computation (using \eqref{eq:projectors_properties} and \eqref{eq:ZF}) 
shows that
\begin{equation}
 \cR(z_1-z_2) \bm{\cA}(z_1) \otimes \bm{\cA}(z_2) = \bm{\cA}(z_2) \otimes \bm{\cA}(z_1).
\end{equation}
Using \eqref{eq:GZ} it is also possible to show that
\begin{equation}
 \bbra{W}\cK(z) \bm{\cA}(-z) = \bbra{W}\bm{\cA}(z) \quad \mbox{and} \quad \overline{\cK}(z)\bm{\cA}(-z)\kket{V} = \bm{\cA}(z)\kket{V}.
\end{equation}

Following the procedure explained in subsection \ref{subsec:stationary_state}, we know that the stationary state of the symmetric generalized exclusion process 
is given by
\begin{equation}
 \steady = \frac{1}{Z_L}\bbra{W} \bm{\cA}(\kappa) \otimes \bm{\cA}(-\kappa) \otimes \bm{\cA}(\kappa) \otimes \cdots \otimes \bm{\cA}(\kappa) \kket{V}.
\end{equation}
We will see below how we can use this algebraic structure to compute physical quantities.

\paragraph{The asymmetric case.}

Similarly to the symmetric case, we introduce the vector
\begin{equation} \label{eq:GASEP_fused_matrix_ansatz}
 \bm{\cA}(z) = Q^{(l)} \bm{A}\left(\frac{z}{t}\right) \otimes \bm{A}\left(zt\right)
\end{equation}
where $\bm{A}(z)$ is the vector, defined in \eqref{eq:ASEP_A}, used to construct the stationary state of the ASEP (with at most one particle per site,
\textit{i.e} for $s=1$).
Writing it out explicitly it has the following expression
\begin{equation} \label{eq:GASEP_A}
 \bm{\cA}(z) = \begin{pmatrix}
                z^2 + z\left(t+\frac{1}{t}\right)\bm{e} + \bm{e^2} \\ 
                t^2 + \frac{1}{t^2} + \left(t+\frac{1}{t}\right)\left(z\bm{d}+\frac{1}{z}\bm{e}\right) + \bm{ed} + \bm{de} \\
                \frac{1}{z^2} + \frac{1}{z}\left(t+\frac{1}{t}\right)\bm{d} + \bm{d^2}
               \end{pmatrix}
\end{equation}
From expression \eqref{eq:GASEP_fused_matrix_ansatz}, a direct computation (using \eqref{eq:projectors_properties_bis} and \eqref{eq:ZF}) 
shows that
\begin{equation}
 \cR\left(\frac{z_1}{z_2}\right) \bm{\cA}(z_1) \otimes \bm{\cA}(z_2) = \bm{\cA}(z_2) \otimes \bm{\cA}(z_1).
\end{equation}
Using \eqref{eq:GZ} it is also possible to show that
\begin{equation}
 \bbra{W}\cK(z) \bm{\cA}\left(\frac{1}{z}\right) = \bbra{W}\bm{\cA}(z) \quad \mbox{and} \quad 
 \overline{\cK}(z)\bm{\cA}\left(\frac{1}{z}\right)\kket{V} = \bm{\cA}(z)\kket{V}.
\end{equation}

Following the procedure explained in subsection \ref{subsec:stationary_state}, we know that the stationary state of the asymmetric generalized exclusion process 
is given by
\begin{equation}
 \steady = \frac{1}{Z_L}\bbra{W} \bm{\cA}(\kappa) \otimes \bm{\cA}\left(\frac{1}{\kappa}\right) \otimes \bm{\cA}(\kappa) \otimes \cdots \otimes \bm{\cA}(\kappa) \kket{V}.
\end{equation}

\subsection{Computation of physical quantities.}

We stress that, despite the apparent complexity of the matrix product solution of the generalized exclusion processes which we introduced, it is 
possible to compute interesting physical quantities.

\paragraph{The symmetric case.}

Once again the first quantity we would like to evaluate is the normalization $Z_L$, which is defined as
$Z_L=\bbra{W}\bm{\cC}(\kappa)\bm{\cC}(-\kappa)\cdots \bm{\cC}(\kappa)\kket{V}$, with $\bm{\cC}(z)=\bra{\sigma}\bm{\cA}(z)$. But looking at the 
expression \eqref{eq:GSSEP_fused_matrix_ansatz} and at the definition of $Q^{(l)}$, it is straightforward to realize that
\begin{eqnarray}
 \bm{\cC}(z) & = & \bra{\sigma}\bm{\cA}(z) \\
 & = & (1,1,1).Q^{(l)} \bm{A}\left(z-\frac{1}{2}\right) \otimes \bm{A}\left(z+\frac{1}{2}\right) \\
 & = & \bm{C}\left(z-\frac{1}{2}\right) \bm{C}\left(z+\frac{1}{2}\right) \\
 & = & (\bm{E}+\bm{D})^2.
\end{eqnarray}
From that fact and the result \eqref{eq:SSEP_normalisation} we deduce directly the following expression of the normalization
\begin{equation} \label{eq:GSSEP_normalisation}
 Z_L = \frac{(a+c)^{2L}(b+d)^{2L}}{(ab-cd)^{2L}} \frac{\Gamma\left(2L+\frac{1}{a+c}+\frac{1}{b+d}\right)}{\Gamma\left(\frac{1}{a+c}+\frac{1}{b+d}\right)}
\end{equation}
which is identical to the normalization of the strict exclusion process (\textit{i.e} with $s=1$) with a system size doubled.

\begin{remark}
 The formula \eqref{eq:GSSEP_normalisation} suggests a connexion with a model with twice the number of sites. The fusion procedure can indeed be 
 interpreted as the 'projection' of a two-lane process onto the generalized exclusion process (with $s=2$). Nevertheless the information about 
 the precise content of the sites in the two-lane process is lost during this projection (we only know after the projection the sum of the number
 of particles of two parallel sites on the two-lane rather than their individual contents). Therefore there does not seem to exist a mapping between 
 the generalized exclusion process and the two-lane process.
\end{remark}

We can also compute the mean particle density at site $i$ (\textit{i.e} the mean number of particles at site $i$). 
For such a purpose the first step is to evaluate, using the explicit expression \eqref{eq:GSSEP_A}, the quantity
\begin{equation}
 (0,1,2).\bm{\cA}(z) = 2z(\bm{E}+\bm{D})+(\bm{E}+\bm{D})\bm{D}+\bm{D}(\bm{E}+\bm{D}).
\end{equation}
Taking into account the averaging over the two steps of the Floquet dynamics (similarly to what we did in the $s=1$ case), it follows that
\begin{equation}
 \langle \tau_i \rangle = 
 \frac{\frac{a}{a+c}\left(2L+\frac{1}{b+d}-2i+\frac{1}{2}\right)+\frac{d}{b+d}\left(2i-\frac{3}{2}+\frac{1}{a+c}\right)}{2L+\frac{1}{a+c}+\frac{1}{b+d}-1}.
\end{equation}
We can also compute the mean particle current. Analyzing carefully all the possible local transitions, we reach, after a long computation,
the expression
\begin{equation}
 \langle J \rangle = 4 \kappa \frac{\bbra{W}(\bm{E}+\bm{D})^{2L-1}\kket{V}}{\bbra{W}(\bm{E}+\bm{D})^{2L}\kket{V}}
 = 4 \kappa \frac{\frac{a}{a+c}-\frac{d}{b+d}}{2L+\frac{1}{a+c}+\frac{1}{b+d}-1}.
\end{equation}

\paragraph{The asymmetric case.}

We mention briefly some computations that can be done in the asymmetric case. Similarly to the $s=1$ case, we will not enter the full details of the 
computations to avoid dealing with the explicit representation of the matrix ansatz algebra (which would require a separate study).
We rather focus on showing that most of the computations are in fact very close to the $s=1$ case.
The first quantity we would like to compute is the normalization $Z_L$, which is defined as
$Z_L=\bbra{W}\bm{\cC}(\kappa)\bm{\cC}(1/\kappa)\cdots \bm{\cC}(\kappa)\kket{V}$, with $\bm{\cC}(z)=\bra{\sigma}\bm{\cA}(z)$. But looking at the 
expression \eqref{eq:GASEP_fused_matrix_ansatz} and at the definition of $Q^{(l)}$, we see that
\begin{eqnarray}
 \bm{\cC}(z) & = & \bra{\sigma}\bm{\cA}(z) \\
 & = & (1,1,1).Q^{(l)} \bm{A}\left(\frac{z}{t}\right) \otimes \bm{A}\left(zt\right) \\
 & = & \bm{C}\left(\frac{z}{t}\right) \bm{C}\left(zt\right) \\
 & = & \left(\frac{z}{t}+\frac{t}{z}+\bm{e}+\bm{d}\right)\left(zt+\frac{1}{zt}+\bm{e}+\bm{d}\right).
\end{eqnarray}
 We thus have
\begin{equation}
 Z_L = \bbra{W}\left(\frac{\kappa}{t}+\frac{t}{\kappa}+\bm{e}+\bm{d}\right)^L\left(\kappa t+\frac{1}{\kappa t}+\bm{e}+\bm{d}\right)^L \kket{V}.
\end{equation}
We can also provide a partial computation for the mean particle current. Analyzing again all the possible local transitions, we end up after 
a long computation to
\begin{equation} \label{eq:GASEP_current}
 \langle J \rangle = \left(\frac{1}{\kappa}-\kappa\right) 
 \frac{\bbra{W}\left(\frac{\kappa}{t}+\frac{t}{\kappa}+\bm{e}+\bm{d}\right)^{L-1}
 \left[2\left(\kappa+\frac{1}{\kappa}\right)+\left(t+\frac{1}{t}\right)(\bm{e}+\bm{d})\right]
 \left(\kappa t+\frac{1}{\kappa t}+\bm{e}+\bm{d}\right)^{L-1} \kket{V}}
 {\bbra{W}\left(\frac{\kappa}{t}+\frac{t}{\kappa}+\bm{e}+\bm{d}\right)^L\left(\kappa t+\frac{1}{\kappa t}+\bm{e}+\bm{d}\right)^L \kket{V}}.
\end{equation}
Once again, using the explicit representation of the matrix ansatz algebra, we should be able to compute the asymptotic behavior of the 
current and draw the exact phase diagram of the model. 
The matrices $\bm{e}$ and $\bm{d}$ are indeed the same as the ones appearing in the solution of the continuous time ASEP and the 
asymptotic behavior of \eqref{eq:GASEP_current} should be derived through the study of contour integral expression similarly to the 
continuous time ASEP (see \cite{MartinRev} for a review).
This will be the subject of a future work.

\section{Conclusion. \label{sec:conclusion}}

There remains a lot of questions to investigate, both from the physical and the mathematical point of view.

The most direct work is to analyse the physical behavior of the generalized exclusion processes which we introduced.
We only gave a very brief flavor of the computation of physical quantities that could be performed using the matrix product solutions.
As already mentioned, we need to perform the computation of the mean particle density and current in the asymmetric case, using the 
explicit representation of the matrices $\bm{e}$ and $\bm{d}$. The asymptotic behavior of the current would yield the phase diagram 
of the model. In particular it would be interesting to study more precisely what are the precise physical effect of i) the discrete time update in
comparison to the continuous time update ii) the small relaxation of the exclusion principle (\textit{i.e} having $s=2$) in comparison to 
the strict exclusion principle ($s=1$).

We are also interested in computing more involved physical quantities such as the large deviation functional of the density profile in the 
stationary state. The starting point would be to write an additivity formula for the stationary weights using the matrix product structure, 
similarly to what was done in the strict exclusion ($s=1$) case (see \cite{DerrReview} for a review). The large deviation functional of the 
density profile has only been computed for a few different models including the SSEP \cite{DLS1}, the ASEP \cite{DLS3}, 
and more recently for a multi-species SSEP \cite{Vanicat17}. The computation of this functional in the generalized exclusion case could give
more insight on the general structure of such functionals.

It would also be interesting to give a hydrodynamic description of the symmetric generalized exclusion process (introduced in 
subsection \ref{subsec:fusion}) in the Macroscopic Fluctuation Theory framework \cite{GianniRevue}. For such a purpose 
a challenging problem would be to compute the diffusion constant and the conductivity associated to the model. Some promising results
have been already obtained in similar but different models \cite{AritaMK,AritaMK2,AritaMK3}. The exact expression of the large deviation functional of 
the density profile could also be of great help to guess or check the expression of the transport coefficients (using the predictions 
of the Macroscopic Theory for the large deviation of the density profile).
All these aforementioned problems will be the object of future works.

On a longer term perspective, we should investigate the construction of higher dimensional generalized exclusion processes, where more
than two particles are allowed on the same site (\textit{i.e} where $s>2$). The fusion procedure can be again applied and interesting results
have already been obtained concerning the exact expression of the fused R-matrix at a particular point \cite{KunibaMMO}. The difficult part of the 
work would be to handle the fusion of the K-matrices (\textit{i.e} deriving closed formulas for the entries). It should be possible to 
express the stationary state of these models using a fused matrix ansatz.

From a more mathematical perspective it would be interesting to explore the connexion of these new models with the theory of 
multivariate orthogonal polynomials. Some recent progress has been made in this direction in the strict exclusion case
\cite{CdGW}. Another challenging task would be to investigate duality relations for these generalized exclusion processes.
The pioneering work \cite{Schutz} led to many promising developments \cite{ImamuraS,GiardinaKRV,BorodinCS,Kuan,ChenDW} 
which provide us with new tools for deriving duality relations.

\section*{Acknowledgments}

It is a pleasure to thank N. Cramp\'{e}, T. Prosen, E. Ragoucy and L. Zadnik for discussions and for their interest in this work.
We are also grateful to C. Finn for his interest and a careful reading of this manuscript. 
We thank the LAPTh, where part of the work has been done, for hospitality and financial support. We acknowledge the 
financial support by the ERC under the Advanced Grant 694544 OMNES.

\end{document}